\documentclass[a4paper,11pt]{article}
\pdfoutput=1

\usepackage{amssymb,amsmath,bm}
\usepackage{a4wide}
\usepackage{color}
\usepackage{slashed}
\usepackage{graphicx}
\usepackage[latin1]{inputenc}
\usepackage{amsfonts}
\usepackage{lscape}
\usepackage{hyperref}
\usepackage{amsthm}

\hypersetup{
    colorlinks=true, 
    linktoc=all,     
    linkcolor=blue,  
    citecolor= black
}

\usepackage{booktabs}
\usepackage{array}
\usepackage{rotating}
\usepackage[numbers, sort&compress]{natbib}
\usepackage{float}
\usepackage{siunitx}
\usepackage[small,bf]{caption}
\newcommand{\GeV}{{\, \rm GeV}}
\newcommand{\TeV}{{\, \rm TeV}}
\newcommand{\eps}{\varepsilon}

\newcommand{\be}{\begin{equation}}
\newcommand{\ee}{\end{equation}}

\newcommand{\abs}[1] {\left| #1 \right|}

\newcommand{\cref}[1]{Chapter~\ref{ch:.#1}}

\newcommand{\beq}{\begin{equation}} 
\newcommand{\eeq}{\end{equation}} 
\newcommand{\ba}{\begin{array}}  
\newcommand{\ea}{\end{array}} 
\newcommand{\bea}{\begin{eqnarray}}  
\newcommand{\eea}{\end{eqnarray} }  
\newcommand{\bal}{\begin{align}}
\newcommand{\eal}{\end{align}}   
\newcommand{\bi}{\begin{itemize}}  
\newcommand{\ei}{\end{itemize}}  
\newcommand{\ben}{\begin{enumerate}}  
\newcommand{\een}{\end{enumerate}}  
\newcommand{\bc}{\begin{center}}
\newcommand{\ec}{\end{center}} 
\newcommand{\bt}{\begin{table}}
\newcommand{\et}{\end{table}}  
\newcommand{\btb}{\begin{tabular}}
\newcommand{\etb}{\end{tabular}}

\begin{document}

\vspace{1cm}
\begin{titlepage}
\vspace*{-1.0truecm}
\begin{flushright}
 CERN-TH-2018-121 \\
 TTP18-019 \\
 \vspace*{2mm}
 \end{flushright}
\vspace{1.8truecm}

\begin{center}
\boldmath

{\Large\textbf{A Realistic U(2) Model of Flavor
}}
\unboldmath
\end{center}

\vspace{0.4truecm}

\begin{center}
{\bf Matthias Linster$^a$ and Robert Ziegler$^{a,b}$}
\vspace{0.4truecm}

{\footnotesize

$^a${\sl Institut f\"ur Theoretische Teilchenphysik, Karlsruhe Institute of Technology, \\ Engesserstra\ss e 7, 76128 Karlsruhe, Germany\vspace{0.2truecm}}

$^b${\sl Theoretical Physics Department, CERN, 1211 Geneva 23, Switzerland \vspace{0.2truecm}}

}
\end{center}

\begin{abstract}
\noindent 
We propose a simple $U(2)$ model of flavor compatible with an $SU(5)$ GUT structure. All hierarchies in fermion masses and mixings arise from powers of two small parameters that control the $U(2)$ breaking. In contrast to previous $U(2)$ models this setup can be realized without supersymmetry and provides an excellent fit to all SM flavor observables including neutrinos. We also consider a variant of this model based on a $D_6 \times U(1)_F$ flavor symmetry, which closely resembles the $U(2)$ structure, but allows for Majorana neutrino masses from the Weinberg operator. Remarkably, in this case one naturally obtains large mixing angles in the lepton sector from small mixing angles in the quark sector. The model also offers a natural option for addressing the Strong CP Problem and Dark Matter by identifying the Goldstone boson of the $U(1)_F$ factor as the QCD axion.
 \end{abstract}

\end{titlepage}

\newpage

\renewcommand{\theequation}{\arabic{section}.\arabic{equation}}


\section{Introduction}


One of the prominent problems of the Standard Model (SM) is the presence of large hierarchies in fermion masses and mixings. Even neglecting neutrino masses, which might have a different origin, the Yukawa couplings span a range from $10^{-6}$ for the electron up to unity for the top quark. Mixing angles in the quark sector are small and hierarchical, while all mixing angles in the lepton sector are sizable. Explaining these hierarchies is referred to as the ``SM Flavor Puzzle" (see e.g. Ref.~\cite{Ferruccio} for a review).

A popular framework to address this problem is in terms of approximate flavor (or horizontal) symmetries. The SM fermions are charged under this symmetry, so that  most of the Yukawa couplings are forbidden in the symmetry limit. The flavor symmetry is spontaneously broken by vacuum expectation values of scalar fields (the so-called flavons), which allows to estimate the Yukawa couplings using a spurion analysis.  Within an effective field theory approach, appropriate powers of flavon insertions are needed to make a given Yukawa operator invariant under the flavor symmetry, suppressed by some large UV cutoff scale. The flavon VEVs are assumed to be slightly below this cutoff scale, so that SM Yukawas arise from powers of these small order parameters. The effective operators have coefficients that are not predicted by the model, but should not be too large or small, in order to explain all hierarchies with the approximate flavor symmetry alone. 

While a plethora of this kind of models have been constructed (see e.g. Ref.~\cite{Ferruccio} and references therein), a particularly simple and interesting class of models is based on a $U(2)$ flavor symmetry~\cite{BarbieriU21, BarbieriU22}. In the original model the flavor quantum numbers are compatible with an  $SO(10)$ GUT structure, and therefore viable only in a supersymmetric (SUSY) context (or more generally in models with at least one additional Higgs field, needed to account for the $m_b/m_t$ hierarchy). Holomorphy together with the $U(2)$ breaking pattern by two spurions then leads to three texture zeros in the quark mass matrices, which imply certain relations between CKM mixing angles and quark masses, in particular $V_{ub}/V_{cb} = \sqrt{m_u/m_c}$. Unfortunately, this prediction is incompatible with the current experimental precision of $V_{ub}$ and $V_{cb}$, and this simple and economic model was ruled out~\cite{RomaninoRoss} with the advent of $B$-factories.   

Therefore modifications of the original model have been proposed in order to modify the model predictions and comply with experimental data. In Ref.~\cite{Dermisek:1999vy} a SUSY $SO(10)$ model with a $D_3 \times U(1)$ flavor symmetry was studied, which mimicked the original $U(2)$ structure with three texture zeros, but is also in conflict with present values of CKM elements. A more recent study has been performed in Ref.~\cite{DGPZ}, which has shown that the problematic relation can be fixed by taking flavor quantum numbers compatible only with an $SU(5)$ GUT structure. This allows the presence of large rotations in the right-handed (RH) down sector that  correct the predictions, as suggested in Ref.~\cite{RomaninoRoss}. Relaxing the $SO(10)$ structure admits to consider also non-supersymmetric models, and in Ref.~\cite{FNZ} such a model was constructed with a charged lepton sector designed to address the (still existing) anomalies in semileptonic $B$-meson decays. This requires to give up also the $SU(5)$ compatibility, but the model can successfully  explain the observed deviations in $R_K$~\cite{RKLHCb} and $R_{K^*}$~\cite{RKstarLHCb} by the tree-level exchange of a $Z^\prime$ boson in the TeV range. In contrast to many $Z^\prime$ models that address the anomalies, the  couplings to fermions are related to the  flavor sector and thus essentially predicted in terms of fermion masses and mixings. 

In this work we build upon the previous studies  in Refs.~\cite{BarbieriU21, DGPZ, FNZ} and propose a simple, non-supersymmetric $U(2)$ model of flavor that is compatible with an $SU(5)$ GUT structure. The problematic relations between CKM mixing angles and quark masses are modified due to large mixing angles in the RH down sector, allowing for an excellent fit to CKM angles and quark and charged lepton masses. All hierarchies arise from powers of two small parameters (roughly of the same order) describing the $U(2)$ breaking pattern.  We also include the neutrino sector, which in this framework can be straightforwardly reproduced by adding three light SM singlets with suitable $U(2)$ quantum numbers and  Dirac masses. The fit to the full SM fermion sector is excellent, and predicts the overall mass scale in the neutrino sector below current cosmological bounds. We further discuss a variant of the $U(2)$ model where the $SU(2)$ factor is replaced by the discrete group $D_6$. The breaking pattern and  the resulting Yukawa matrices closely resemble the $SU(2)$ case. The only difference is a flipped sign in the 1-2 entry of the mass matrices, which has no effect in the quark and charged lepton sector, but allows to obtain Majorana neutrinos masses from the Weinberg operator. In contrast to the Dirac case the parametric flavor suppression of the neutrino mass matrix is fixed purely by charged lepton charges. Remarkably, this matrix is automatically anarchical, and therefore allows for an excellent fit to neutrino data, again predicting the overall neutrino mass scale in about the same range as in the Dirac case. 

Finally we discuss the fate of the $U(1) \subset U(2)$ Goldstone boson, which naturally plays the role of the QCD axion and has (flavor-violating) couplings to fermions that are predicted by the flavor model, in the spirit of Refs.~\cite{Wilczek, Axiflavon, Japanese}. In contrast to single $U(1)$ flavor models, here the additional $SU(2)$ flavor symmetry protects flavor-violating couplings to light generations (much as in SUSY U(2) models~\cite{DGPZ, FNZ}),  so that the resulting axion is mainly constrained by astrophysics and not by precision flavor observables. It is well-known that the axion can be an excellent Dark Matter (DM) candidate for large ranges of the $U(1)$ breaking scale, which here is directly connected to the UV cutoff of the flavor model.  In this way the model offers a natural solution for the strong CP problem and the origin of DM.

This paper is organized as follows. In Section 2 we define the $U(2)$ flavor model  and discuss the structure of the quark and charged lepton sector before addressing the (Dirac) neutrino sector. We then consider a $D_6 \times U(1)$ model in Section 3, which closely follows the $U(2)$ structure and allows to obtain Majorana neutrino masses from the Weinberg operator. In Section 4 we address the Strong CP Problem and Dark Matter within this framework, interpreting the Goldstone boson of the $U(1)$ factor as the QCD axion. We finally conclude in Section 5. In three Appendices we provide more details on the group theoretical structure of $D_3$ and $D_6 \simeq D_3 \times Z_2$, include more details about the numerical fit, and discuss an explicit example of the scalar potential generating the flavon VEVs in the $D_6 \times U(1)$ model.

\section{A Realistic $U(2)$ Model of Flavor}
\setcounter{equation}{0}

In this section we define our framework and show how hierarchies in the quark and charged lepton sector arise from the $U(2)$ flavor symmetry. After discussing the analytical relations between CKM elements and quarks masses, we perform a numerical fit to masses and mixings. We then address the neutrino sector in the context of Dirac neutrinos and include it in the numerical fit.  We conclude this section with a general discussion of the flavor structure of neutrino masses, motivating the $D_6 \times U(1)$ flavor model in the next section.
 
\subsection{Quark and Charged Lepton Sector}

We consider an extension of the SM with a global flavor symmetry group $U(2)_F$. Locally this group is isomorphic to $SU(2)_F \times U(1)_F$, under which SM fermions are charged. This symmetry group is assumed to be broken slightly below a UV scale $\Lambda$, which sets the relevant mass scale for additional dynamics. We also assume that the scale $\Lambda$ is large enough to safely neglect the impact of these new degrees of freedom on phenomenology. Thus, we simply work with  an effective theory with cut-off scale $\Lambda$ that only involves SM fields and spurions that parametrize the breaking of $SU(2)_F \times U(1)_F$.

The SM fermions have $U(2)_F$ quantum numbers that are compatible with an $SU(5)$ GUT structure, i.e. they are specified by the quantum number of the two $SU(5)$ representations ${\bf 10} = Q,U,E$ and   ${\bf \overline{5}} = L,E$.  The first two generations transform as a doublet under $SU(2)_F$, the third generation is an  $SU(2)_F$ singlet and the Higgs field is a singlet under both $SU(2)_F$ and $U(1)_F$. Thus, the $U(1)_F$ quantum numbers of the SM fermions are specified by four charges \{$X_{{10}_a}, X_{\overline{5}_a}, X_{{10}_3},  X_{{\overline{5}}_3}$\} for \{${\bf 10}_a,  {\bf \overline{5}}_a, {\bf 10}_3, {\bf \overline{5}}_3$\} with $a=1,2$. It turns out that a successful fit to the observed fermion masses and mixings can be achieved for the following simple choice for $U(1)_F$ charges:
\begin{align}
X_{{10}_3} & = 0 \, , & X_{{10}_a} & = X_{{\overline{5}}_a} = X_{{\overline{5}}_3} = 1 \, .
\end{align}
The breaking of the flavor symmetry  is described by two scalar spurions $\phi$ and $\chi$, which transform under $U(2)_F$ as  $\phi$ = ${\bf 2}_{-1}$ and $\chi$= ${\bf 1}_{-1}$.  
These fields acquire the following vacuum expectation values (VEVs):
\begin{align}
\langle \phi \rangle & =  \begin{pmatrix} \eps_\phi \Lambda \\ 0 \end{pmatrix} \, , &
 \langle \chi \rangle & = \eps_\chi \Lambda \, ,
\end{align}
where we will take $\eps_{\phi} \sim \eps_{\chi} \sim {\cal O}(0.01)$.  In Table~\ref{tab:charges} we summarize the field content and the transformation properties under the flavor group. 
\begin{table}[h]
\centering
\begin{tabular}{cccccccc}
\toprule
& ${\bf 10}_a$ & ${\bf \overline{5}}_a$ &${\bf 10}_3$ & ${\bf \overline{5}}_3$  &  $H$ & $\phi_a$ & $\chi$ \\
\midrule
$SU(2)_F$ & ${\bf 2}$ & ${\bf 2}$  & ${\bf 1}$ & ${\bf 1}$ & ${\bf 1}$ & ${\bf 2}$ & ${\bf 1}$ \\    
$U(1)_F$ & $1$ & $1$ & $0$ & $1$ & $0$ & $-1$ & $-1$    \\
\bottomrule
\end{tabular}
\caption{The field content and $U(2)_F$ quantum numbers. \label{tab:charges}}
\end{table}
As the fermions are charged under $U(2)_F$, Yukawa couplings require additional spurion insertions in order to be $U(2)_F$-invariant. 
This leads to non-renormalizable interactions  suppressed by appropriate powers of $\Lambda$. For example,  the resulting Lagrangian in the up-sector, at leading order in  $\eps_{\phi, \chi}$, is given by
\begin{align}
\label{Lu}
{\cal L}_u & = \frac{\lambda^u_{11}}{\Lambda^6} \chi^4  (\phi^*_a Q_a) (\phi^*_b U_b) H  + \frac{\lambda^u_{12}}{\Lambda^2} \chi^2  \epsilon_{ab} Q_a U_b H + \frac{\lambda^u_{13}}{\Lambda^3} \chi^2   (\phi^*_a Q_a)  U_3 H  \nonumber \\
& +  \frac{\lambda^u_{22}}{\Lambda^2}    (\epsilon_{ab} \phi_a Q_b)   (\epsilon_{cd} \phi_c U_d)   H + \frac{\lambda^u_{23}}{\Lambda}    (\epsilon_{ab} \phi_a Q_b)  U_3 H +   \frac{\lambda^u_{31}}{\Lambda^3} \chi^2   Q_3 (\phi^*_a U_a)   H \nonumber \\
& + \frac{\lambda^u_{32}}{\Lambda}  Q_3  (\epsilon_{ab} \phi_a U_b)  H + \lambda^u_{33} Q_3 U_3 H \, , 
\end{align}
and similar in the down and charged lepton sector. After inserting the spurion VEVs the cutoff dependence drops out, and Yukawa hierarchies arise from powers of the small parameters $\eps_{\phi, \chi}$. In this way we get for the up-, down- and charged lepton Yukawa matrices (defined as ${\cal L}_{\rm yuk} = Q^T Y_u U H + \cdots  $)  the result 
\begin{gather}
\label{eq:yf}
Y_u  \approx
\begin{pmatrix}
\lambda_{11}^u \eps_\phi^2  \eps_\chi^4 & \lambda_{12}^u \eps_\chi^2  & \lambda_{13}^u \eps_\phi  \eps_\chi^2  \\ 
- \lambda_{12}^u \eps_\chi^2 & \lambda_{22}^u \eps_\phi^2  & \lambda_{23}^u \eps_\phi \\
\lambda_{31}^u \eps_\phi  \eps_\chi^2 & \lambda_{32}^u \eps_\phi & \lambda_{33}^u
\end{pmatrix} \, , \qquad
Y_d  \approx
\begin{pmatrix}
\lambda_{11}^d \eps_\phi^2  \eps_\chi^4 & \lambda_{12}^d \eps_\chi^2  & \lambda_{13}^d \eps_\phi  \eps_\chi^3  \\ 
- \lambda_{12}^d \eps_\chi^2 & \lambda_{22}^d \eps_\phi^2  & \lambda_{23}^d \eps_\phi \eps_\chi \\
\lambda_{31}^d \eps_\phi  \eps_\chi^2 & \lambda_{32}^d \eps_\phi & \lambda_{33}^d \eps_\chi 
\end{pmatrix} \, , \\
Y_e  \approx
\begin{pmatrix}
\lambda_{11}^e \eps_\phi^2  \eps_\chi^4 & \lambda_{12}^e \eps_\chi^2  & \lambda_{13}^e \eps_\phi  \eps_\chi^2  \nonumber \\ 
- \lambda_{12}^e \eps_\chi^2 & \lambda_{22}^e \eps_\phi^2  & \lambda_{23}^e \eps_\phi  \\
\lambda_{31}^e \eps_\phi  \eps_\chi^3 & \lambda_{32}^e \eps_\phi \eps_\chi & \lambda_{33}^e \eps_\chi 
\end{pmatrix} \, ,
\end{gather}
where $\lambda_{ij}^f$ are (in general complex) ${\cal O}(1)$ coefficients and we have kept only the leading contributions in $\eps_{\phi, \chi}$. Note that, in contrast to the supersymmetric $U(2)$ model in Ref.~\cite{DGPZ}, there are no holomorphy constraints, which leads to a more general Yukawa pattern. 

One can show that the $\lambda_{11}, \lambda_{13}, \lambda_{31}$ entries give only subleading corrections to quark masses and mixings, which are relatively suppressed by at least $\eps_\phi^2$. Thus, effectively, three texture zeros appear in the Yukawa matrix, much as in the supersymmetric models~\cite{DGPZ}, and to good approximation we obtain the Yukawa couplings 
\begin{gather}
\label{eq:yfapp}
Y_u  \approx
\begin{pmatrix}
0 & \lambda_{12}^u \eps_\chi^2  & 0  \\ 
- \lambda_{12}^u \eps_\chi^2 & \lambda_{22}^u \eps_\phi^2  & \lambda_{23}^u \eps_\phi \\
0 & \lambda_{32}^u \eps_\phi & \lambda_{33}^u
\end{pmatrix} \, , \qquad
Y_d  \approx
\begin{pmatrix}
0 & \lambda_{12}^d \eps_\chi^2  &0  \\ 
- \lambda_{12}^d \eps_\chi^2 & \lambda_{22}^d \eps_\phi^2  & \lambda_{23}^d \eps_\phi \eps_\chi \\
0 & \lambda_{32}^d \eps_\phi & \lambda_{33}^d \eps_\chi 
\end{pmatrix} \, , \nonumber \\
Y_e  \approx
\begin{pmatrix}
0  & \lambda_{12}^e \eps_\chi^2  & 0  \\ 
- \lambda_{12}^e \eps_\chi^2 & \lambda_{22}^e \eps_\phi^2  & \lambda_{23}^e \eps_\phi  \\
0 & \lambda_{32}^e \eps_\phi \eps_\chi & \lambda_{33}^e \eps_\chi 
\end{pmatrix} \, .
\end{gather}
Because of the hierarchical structure and the presence of the texture zeros, it is possible to analytically derive  some approximate results for the singular values and the rotations to the mass basis~\cite{FNZ}. One can also perturbatively diagonalize the Yukawa matrices, and obtain the following estimates for singular values and CKM matrix elements (neglecting ${\cal O}(1)$ coefficients):
\begin{align}
y_u & \sim  \eps_\chi^4/\eps_\phi^2 \, , & y_d & \sim y_e \sim  \eps_\chi^4/\eps_\phi^2  \, , & V_{ub} & \sim \eps_\chi^2/\eps_\phi \, , \nonumber \\
y_c & \sim \eps_\phi^2   \, , & y_s & \sim y_\mu \sim \eps_\phi^2 \eps_\chi/ \sqrt{\eps_\phi^2 + \eps_\chi^2} \, , &  V_{cb} & \sim \eps_\phi \, , \nonumber \\
 y_t & \sim 1  \, , & y_b & \sim y_\tau \sim  \sqrt{\eps_\phi^2 + \eps_\chi^2}  \,, & V_{us} & \sim  \eps_\chi^2/\eps_\phi^2 \, .
\label{QCLanalytic}
\end{align}
These expressions can be compared to the ($1 \sigma$) ranges for fermion mass ratios and CKM elements, taken for definiteness at 10 TeV 
\begin{align}
\frac{m_u}{m_t} & \approx \lambda^{(7.1 \div 7.7)} \, , & \frac{m_d}{m_b} & \approx \lambda^{(4.2 \div 4.4)} \, , & \frac{m_e}{m_\tau} & \approx \lambda^{5.1} \, , & V_{ub} & 
\approx \lambda^3 \nonumber \\
\frac{m_c}{m_t} & \approx \lambda^{3.5} \, , 
& \frac{m_s}{m_b} & \approx \lambda^{(2.4 \div 2.5)} \, ,  & \frac{m_\mu}{m_\tau} & \approx \lambda^{1.8} \, , & V_{cb} & \approx \lambda^2 \, , 
\end{align}
where $\lambda = 0.2 \approx V_{us}$ and  $y_b (10 \TeV) \approx \lambda^{2.7}$, $y_\tau (10 \TeV) \approx \lambda^{2.8}$.  Within roughly a factor $\lambda$, all hierarchies can be reproduced taking  
\begin{align}
\eps_\phi & \sim V_{cb} \sim \lambda^2  \, , & \eps_\chi &  \sim \lambda^{2 \div 3}  \, ,
\label{epsestimate}
\end{align} 
and  therefore a good fit to masses and mixings can be expected with  input parameters $\lambda_{ij}^f$ that are indeed ${\cal O}(1)$. Moreover,  it is clear that there must be four relations in each fermion sector between the 3 singular values and the 3+3 rotation angles. For real $h_{ij}^f$ it is straightforward to work out these predictions exactly~\cite{FNZ} and expand the result in ratios of the hierarchical eigenvalues. One can then relate the 1-2 and 1-3 rotations in the left- and right-handed sectors to the 2-3 rotations and the eigenvalues. With the convention 
\begin{align}
Y & = V_L Y_{\rm diag} V_R^\dagger \, , & V_L & = V^L_{13} V_{12}^L V_{23}^L \, , & V_R & = V^R_{13} V_{12}^R V_{23}^R \, ,   
\end{align}  
where $V_{ij}$ are orthogonal rotation matrices in the $i$-$j$ plane that are parametrized by the angles $s_{ij} \equiv \sin \theta_{ij}$, one obtains up to percent corrections 
\begin{align}
\label{eq:approxrotangles}
s^{Lu}_{12}  & \approx -s^{Ru}_{12} \approx  \sqrt{\frac{m_u}{m_c}}  \, , &
s^{Lu}_{13} & \approx - s^{Lu}_{23} s_{12}^{Lu} \, , & s^{Ru}_{13} & \approx    s^{Ru}_{23} s^{Lu}_{12} 
\, , \nonumber \\
s^{Ld}_{12}  & \approx -s^{Rd}_{12}  \approx  \sqrt{\frac{m_d}{m_s}}   \sqrt{c_{23}^{Rd}}  \, , &
s^{Ld}_{13} & \approx  - s^{Ld}_{23} s^{Ld}_{12}  \left( 1 - \frac{s^{Rd}_{23}}{c^{Rd}_{23} s^{Ld}_{23}} \frac{m_s}{m_b}\right) \, , & s^{Rd}_{13} & \approx     \frac{s^{Rd}_{23}}{c_{23}^{Rd}} s^{Ld}_{12} \, , \nonumber \\
s^{Re}_{12}  & \approx -s^{Le}_{12}  \approx  \sqrt{\frac{m_e}{m_\mu}}   \sqrt{c_{23}^{Le}}  \, , &
s^{Re}_{13} & \approx  - s^{Re}_{23} s^{Re}_{12}  \left( 1 - \frac{s^{Le}_{23}}{c^{Le}_{23} s^{Re}_{23}} \frac{m_\mu}{m_\tau}\right) \, , & s^{Le}_{13} & \approx     \frac{s^{Le}_{23}}{c_{23}^{Le}} s^{Re}_{12} \, ,
\end{align}
where  2-3 rotations angles are large in the RH down and LH charged lepton sector, and CKM-like in all other sectors
\begin{align}
s_{23}^{Rd} & \sim s_{23}^{Le} \sim 1 \, , & s_{23}^{Lu} & \sim s_{23}^{Ru} \sim s_{23}^{Ld} \sim  s_{23}^{Re} \sim V_{cb} \, .
\end{align}
One therefore obtains for the CKM elements (in our conventions $V_{\rm CKM} = V^{uT}_L V^{d*}_L $) the predictions 
\begin{align}
|V_{ub}| & \approx  \left| \sqrt{\frac{m_u}{m_c}}  |V_{cb}| -  e^{i \phi_1} \sqrt{\frac{m_d}{m_s}} \sqrt{c_{23}^{Rd}}  \frac{s^{Rd}_{23}}{c^{Rd}_{23} } \frac{m_s}{m_b} \right| \, , \quad |V_{td}|  \approx  \sqrt{\frac{m_d}{m_s} } \sqrt{c_{23}^{Rd}} \left|  |V_{cb}| -  e^{i \phi_2}  \frac{s^{Rd}_{23}}{c^{Rd}_{23} } \frac{m_s}{m_b} \right| \, , \nonumber \\
|V_{us}| & \approx |s_{12}^{Ld} - s_{12}^{Lu}| \approx \left| \sqrt{\frac{m_d}{m_s}} \sqrt{c_{23}^{Rd}} - e^{i (\phi_2 - \phi_1)} \sqrt{\frac{m_u}{m_c}} \right| \, , \quad
|V_{cb}|  \approx |V_{ts}| \approx |s_{23}^{Ld} - s_{23}^{Lu}| \, ,
\label{CKMpred}
\end{align}
where we included also relative phases $\phi_{1,2}$, see Ref.~\cite{DGPZ} for details. In the original $U(2)$ models in Ref.~\cite{BarbieriU21, BarbieriU22},  the rotation angle in 2-3 RH down sector $s_{23}^{Rd}$  was taken to be of the order of the other 2-3 rotation angles,  $s_{23}^{Rd}  \sim V_{cb}$. From the above equations, this directly leads to the accurate prediction $ |V_{ub}/V_{cb}|   \approx  \sqrt{m_u/m_c}$ which deviates from experimental data by more than $3 \sigma$. 
This is the reason why here this angle is taken to be large,  $s_{23}^{Rd} \sim c_{23}^{Rd}  \sim 1/\sqrt{2} $,  which then allows to obtain an excellent fit to CKM angles as we demonstrate in the next section (see also Refs.~\cite{RomaninoRoss, DGPZ, FNZ}). 

\subsection{Fit to Quark and Charged Lepton Sector}

We now perform a numerical fit to the model parameter set $\{ \lambda_{ij}^{u,d,e}, \eps_\phi, \eps_\chi\}$. For simplicity, we restrict to real $\lambda_{ij}^{u,d,e}$ and demonstrate later on that the CKM phase can be obtained by taking a complex parameter $\lambda^u_{33}$.  The experimental input parameters are therefore the quark and charged lepton masses and  the CKM mixing angles. For concreteness we take them in the $\overline{{\rm MS}}$ scheme at 10 TeV from Ref.~\cite{AntuschMaurer}, with a symmetrized $1 \sigma$ error taken to be the larger one. All input parameters are summarized in Table~\ref{input}.
\begin{table}[H]%
	\centering
		\begin{tabular}{cc}%
		\toprule%
		Quantity & Value\\%
		\midrule%
		$y_u$ & \num{5.7 \pm 2.3 e-6}\\%
		$y_d$ & \num{1.223 \pm 0.18 e-5}\\%
		$y_s$ & \num{2.42 \pm 0.13 e-4}\\%
		$y_c$ & \num{2.776 \pm 0.088 e-3}\\%
		$y_b$ & \num{1.224 \pm 0.013 e-2}\\%
		$y_t$ & \num{0.7894 \pm 0.0092}\\%
		$y_e$ & \num{2.8782 \pm 0.0042 e-6}\\%
		$y_\mu$ & \num{6.0761 \pm 0.0088 e-4}\\%
		$y_\tau$ & \num{1.0329 \pm 0.0015 e-2}\\%
		\midrule%
		$\theta_{12}$ & \num{0.22736 \pm 0.00072}\\%
		$\theta_{23}$ & \num{4.364 \pm 0.067 e-2}\\%
		$\theta_{13}$ & \num{3.77 \pm 0.14 e-3}\\%
		\bottomrule%
	\end{tabular}%
	\label{tab:fitParametersQuarkAndChargedLeptons}
	\caption{\label{input} Input values of quark and charged lepton Yukawas and quark mixing angles at 10 TeV taken from Ref.~\cite{AntuschMaurer}.}%
\end{table}%
The quality of the fit with a given model parameter set $\{ \lambda_{ij}^{u,d,e}, \eps_\phi, \eps_\chi\}$ is measured by two functions $\chi^2$ and $\chi^2_{{\cal O}(1)}$. The first quantity is the usual $\chi^2$ that indicates how well the experimental input values are reproduced by the fit. It is obtained by plugging the model parameters into the Yukawa matrices in Eq.~\eqref{eq:yfapp} and calculating numerically the singular values $y_{q,l}$ and the CKM mixing angles $\theta_{ij}$ in the PDG parametrization.  These values are used with the experimental input above to obtain $\chi^2$ defined as 
\begin{equation}%
	\label{eq:chi2FitQuarkChargedLeptonSector}
	\chi^2 = \sum_{q = u,d,s,c,b,t} \frac{(y_{q}-y_{q,{\rm exp}})^2}{({\sigma y_{q,{\rm exp}}})^2} + \sum_{\ell = e,\mu,\tau} \frac{(y_{\ell}-y_{\ell,{\rm exp}})^2}{({\sigma y_{\ell,{\rm exp}}})^2} + \sum_{(ij) = (12),(13),(23)} \frac{(\theta_{ij}-\theta_{ij,{\rm exp}})^2}{({\sigma \theta_{ij,{\rm exp}}})^2} \, .
\end{equation}%
In order to explain Yukawa hierarchies solely by $U(2)_F$ breaking, the parameters $\lambda^{u,d,e}_{ij}$ should be ${\cal O}(1)$. The meaning of this requirement is somewhat fuzzy, and here we choose to quantify it by introducing a measure $\chi_{{\cal O}(1)}^2$ defined as
\begin{equation}%
\label{eq:chiO1}
	\chi_{{\cal O}(1)}^2 = \sum_{\lambda_{ij}^p} \frac{\left(\log(|
\lambda_{ij}^p|)\right)^2}{2 \cdot 0.55^2} \, ,
\end{equation}%
where $i,j = 1,2,3$ and $p = u,d,e$. This corresponds to the assumption that  the $\lambda_{ij}^{u,d,e}$ are distributed according to a log-normal distribution with mean $1$ and standard deviation $\sigma = 0.55$, i.e. the absolute values $\lambda_{ij}^{u,d,e}$ lie with a probability of \SI{95}{\%} within the interval $[1/3,3]$. For example, the contribution to $\chi^2_{{\cal O}(1)}$ of a single parameter $\lambda  = \{ 3, 5, 7, 10 , 50 ,100\}$ (or the inverse) is $\Delta \chi^2_{{\cal O}(1)} = \{ 2, 4, 6, 9, 25, 35\}$. We consider a fit satisfactory as long as $\chi^2_{{\cal O}(1)}   \le \# {\rm pars}$, and there are 5 parameters for each fermion sector.   As the best fit we choose the one that minimizes both $\chi^2$ and $\chi_{{\cal O}(1)}^2$. 

In Table~\ref{tab:bestFitsQuarkAndChargedLeptons} we show our fit results, where we display the values of the small parameters $\eps_\phi, \eps_\chi$ and indicate separately the two fit measures $\chi^2$, $\chi_{{\cal O}(1)}^2$ as defined above, along with the smallest and largest $|\lambda_{ij}^{u,d,e}|$. For the fit QL1$_{\mathbb R}$ we have minimized $\chi^2 + \chi^2_{{\cal O}(1)}$, while for QL2$_{\mathbb R}$ we have minimized $\chi^2_{{\cal O}(1)}$ while keeping $\chi^2 \le \# {\rm obs} = 12$. For illustrative purposes we also show a fit that minimizes just $\chi^2$ (QL3$_{\mathbb R}$). 
\begin{table}[H]
	\centering
	\begin{tabular}{|c||cccccc|}
	\hline
		Fit & {$\varepsilon_\phi$} & {$\varepsilon_\chi$} & {min $|\lambda_{ij}^{u,d,\ell}|$} & {max $|\lambda_{ij}^{u,d,\ell}|$} & {$\chi^2$} & {$\chi^2_{{\cal O}(1)}$}\\
		\hline
		\hline
		QL1$_{\mathbb R}$ & 0.019 & 0.008 & 1/3.1 & 2.7 & 1.7 & 7.8\\
		QL2$_{\mathbb R}$ & 0.023 & 0.008 & 1/2.7 & 2.8 & 12 & 5.4\\	
		\hline
		{\small QL3$_{\mathbb R}$} & {\small 0.065 }& {\small 0.011 }& {\small 1/9.1 }& {\small 6.9 }& {\small 0 }& {\small 35}\\
		\hline
	\end{tabular}
	\caption{Best fits in the quark and charged lepton sector.
	\label{tab:bestFitsQuarkAndChargedLeptons}}
\end{table}%
\noindent Indeed there are enough free parameters to obtain a perfect fit to observables, however one needs $\chi^2_{{\cal O}(1)}$ as large as 35 and ${\cal O}(1)$ parameters as small as $\approx 1/9$, so this fit should be discarded according to our quality requirement $\chi^2_{{\cal O}(1)}  < 15$. The best fits are QL1$_{\mathbb R}$ and QL2$_{\mathbb R}$ with ${\cal O}(1)$ parameters between 1/3 and 3, which feature values of $\eps_{\phi}, \eps_\chi$ that are indeed of the naive size estimated in Eq.~\eqref{epsestimate}. 

Finally we demonstrate that the CKM phase $\delta_\mathrm{CP}$ can be easily included. For simplicity we restrict to the case where only the 33 entry in the up-quark Yukawa matrix is complex, i.e. $\lambda^u_{33} \to \lambda^u_{33}  e^{i \delta_{33}} $. In a realistic setup where all Yukawas have phases, the fit can only get better. In the $\chi^2$ measure in Eq.~\eqref{eq:chi2FitQuarkChargedLeptonSector} we now include the CP phase of the CKM matrix, with the experimental value taken from Ref.~\cite{AntuschMaurer}
\begin{equation*}
	\delta_\mathrm{CP, exp} = 1.208 \pm 0.054 \, .
\end{equation*}
Including $\delta_{33}$ leads to even better fits (QL1 and QL2), which we show in Table~\ref{tab:bestFitsQuarkAndChargedLeptonsCP}. This demonstrates that an excellent fit for  quark and charged lepton sector, including the CKM phase, can be obtained with all ${\cal O}(1)$ parameters lying between $1/2.8$ and \num{2.1}. 
\begin{table}[H]
	\centering
	\begin{tabular}{|c||cccccc|}
		\hline
		Fit & {$\varepsilon_\phi$} & {$\varepsilon_\chi$} & {min $|\lambda_{ij}^{u,d,\ell}|$} & {max $|\lambda_{ij}^{u,d,\ell}|$} & {$\chi^2$} & {$\chi^2_{{\cal O}(1)}$}\\
		\hline
		\hline
		QL1 & 0.025 & 0.009 & 1/2.9 & 2.1 & 0.6 & 5.8\\
		QL2 & 0.024 & 0.008 & 1/2.8 & 1.9 & 13 & 4.8\\ 
		\hline
	\end{tabular}
	\caption{Best fits in the quark and charged lepton sector including the CKM phase. 
	\label{tab:bestFitsQuarkAndChargedLeptonsCP}}
\end{table}
\subsection{Neutrino Sector}
In the neutrino sector we have to distinguish whether neutrinos are Dirac or Majorana. We begin with the discussion of the Dirac scenario, since the Majorana case in the $U(2)_F$ model is strongly disfavored as we will discuss below. To this extent we introduce SM singlets $N_a, N_3$ with $U(1)_F$ charges $X^N_a$ and $X^N_3$, where  $N_a$ transforms as a doublet of $SU(2)_F$ and $N_3$ as a singlet. The Lagrangian then allows for a Yukawa coupling ${\cal L}_\nu = L^T Y_\nu N H$ (we assume that the Majorana mass term is forbidden,   e.g. by exact lepton number conservation). As in the charged lepton sector, one can obtain its  structure from a spurion analysis as
\begin{align}%
	Y_\nu & =\begin{pmatrix}
		\lambda_{11}^\nu \eps_\phi^2 \eps_\chi^{\abs{3+X_a^N}} & \lambda_{12}^\nu \eps_\chi^ {\abs{1+X_a^N}} & \lambda_{13}^\nu \eps_\phi \eps_\chi^{\abs{2+X_3^N}}\\%
		-\lambda_{12}^\nu \eps_\chi^{\abs{1+X_a^N}} & \lambda_{22}^\nu \eps_\phi^2 \eps_\chi^{\abs{X_a^N-1}} & \lambda_{23}^\nu \eps_\phi \eps_\chi^{\abs{X_3^N}}\\%
		\lambda_{31}^\nu \eps_\phi \eps_\chi^{\abs{2+X_a^N}} & \lambda_{32}^\nu \eps_\phi \eps_\chi^{\abs{X_a^N}} & \lambda_{33}^\nu \eps_\chi^{\abs{1+X_3^N}}
\end{pmatrix} \, . \label{eq:ynu}
\end{align}%
It is clear that in order to obtain sub-eV neutrinos one needs large $U(1)_F$ charges $X^N_{a,3} > 1$, so that one can drop the absolute values in Eq.~\eqref{eq:ynu}. In this case the contributions from the $(11),(13),(31)$ entries to masses and mixings are again sub-leading, and we can drop them as in the previous section and are left with the Dirac neutrino mass matrix
\begin{align}
m^D_\nu & \approx v \begin{pmatrix}
		0 & \lambda_{12}^\nu \eps_\chi^ {1+X_a^N} & 0 \\%
		-\lambda_{12}^\nu \eps_\chi^{1+X_a^N} & \lambda_{22}^\nu \eps_\phi^2 \eps_\chi^{X_a^N-1} & \lambda_{23}^\nu \eps_\phi \eps_\chi^{X_3^N}\\%
		0 & \lambda_{32}^\nu \eps_\phi \eps_\chi^{X_a^N} & \lambda_{33}^\nu \eps_\chi^{1+X_3^N}
\end{pmatrix} \, .
\end{align}
It is well-known that an anarchical neutrino mass matrix can give a good fit to neutrino observables, which can be achieved taking $X^N_a = X^N_3$ (since $\eps_\chi \sim \eps_\phi$), giving
\begin{align}
m^D_\nu & \approx v \, \eps_\chi^{X_a^N-1} \begin{pmatrix}
		0 & \lambda_{12}^\nu \eps_\chi^ 2 & 0 \\%
		-\lambda_{12}^\nu \eps_\chi^{2} & \lambda_{22}^\nu \eps_\phi^2 & \lambda_{23}^\nu \eps_\phi \eps_\chi \\%
		0 & \lambda_{32}^\nu \eps_\phi \eps_\chi & \lambda_{33}^\nu \eps_\chi^{2}
\end{pmatrix} \, .
\label{mnuD}
\end{align}
In order to obtain an overall neutrino mass scale $\lesssim 0.1 \, {\rm eV}$, one needs $X^N_3 \gtrsim 5$, so that tiny neutrino masses arise from somewhat large $U(1)_F$ charges and the smallness of the $U(2)_F$ breaking parameters, $\eps_{\chi,\phi} \sim 0.01$.

These considerations are confirmed by a numerical fit, for which we proceed as in the previous section, now including the neutrino sector. For the input values for normal (NO) and inverted mass ordering (IO), we use the neutrino mass differences and PMNS mixing angles from the global NuFIT 3.2 (2018) in Refs.~\cite{SchwetzPaper,SchwetzWebsite}, which are summarized in Table~\ref{neutrinoinput}.  
\begin{table}[H]
	\centering
		\begin{tabular}{cc}
		\multicolumn{2}{c}{\bf Normal Ordering (NO)} \\
		\toprule 
		Quantity & Value\\
		\midrule
		$\Delta m_{21}^2$ & \num{7.4 \pm 0.21 e-5}\\
		$\Delta m_{31}^2$ & \num{2.494 \pm 0.033 e-3}\\
		\midrule
		$\sin^2 \theta_{12}$ & \num{0.307 \pm 0.013}\\
		$\sin^2 \theta_{13}$ & \num{0.02206 \pm 0.00075}\\
		$\sin^2 \theta_{23}$ & \num{0.538 \pm 0.069}\\
		\bottomrule
	\end{tabular}
	\qquad \qquad 
		\begin{tabular}{cc}
		\multicolumn{2}{c}{\bf Inverted Ordering (IO) } \\
		\toprule 
		Quantity & Value\\
		\midrule
		$\Delta m_{21}^2$ & \num{7.4 \pm 0.21 e-5}\\
		$\Delta m_{32}^2$ & \num{-2.465 \pm 0.032 e-3}\\
		\midrule
		$\sin^2 \theta_{12}$ & \num{0.307 \pm 0.013}\\
		$\sin^2 \theta_{13}$ & \num{0.02227 \pm 0.00074}\\
		$\sin^2 \theta_{23}$ & \num{0.554 \pm 0.0033}\\
		\bottomrule
	\end{tabular}
	\caption{\label{neutrinoinput} Experimental values of neutrino mass differences and PMNS mixing angles for normal (NO) and inverted hierarchy (IO), taken  from NuFIT 3.2 (2018)~\cite{SchwetzPaper, SchwetzWebsite}.}
\end{table}
We then plug the neutrino model parameters $\lambda^\nu_{ij}$ for fixed charges $X^N_a, X^N_3$ into the Yukawa matrices in Eq.~\eqref{mnuD} and calculate numerically the singular values and the PMNS mixing angles $\theta_{ij}$ in the standard parametrization. To the $\chi^2$ defined in Eq.~\eqref{eq:chi2FitQuarkChargedLeptonSector} we add the corresponding expression $\chi^2_\nu$ in the neutrino sector~\footnote{For the angle $\sin^2 \theta_{23}$ we actually use the full $\chi^2$ function provided by the NuFIT collaboration instead of assuming the Gaussian error in Table~\ref{neutrinoinput}.}
 \begin{equation}%
	\label{eq:chi2NeutrinoSector}
	\chi^2_\nu = \sum_{(ij) = 21,31/32} \frac{(\Delta m^2_{ij}-\Delta m^2_{ij,{\rm exp}})^2}{({\sigma \Delta m^2_{ij,{\rm exp}}})^2}  + \sum_{(ij) = (12),(13),(23)} \frac{(\sin^2 \theta_{ij}-\sin^2 \theta_{ij,{\rm exp}})^2}{({\sigma \sin^2 \theta_{ij,{\rm exp}}})^2} \, ,
\end{equation}%
and similarly we include the coefficients $\lambda_{ij}^\nu$ in the measure $\chi^2_{{\cal O}(1)}$ defined in Eq.~\eqref{eq:chiO1}. We then perform a simultaneous fit to quark, charged lepton and neutrino sector including a phase in $\lambda^u_{33}$ as discussed in the last section (for simplicity we omit phases in the neutrino sector, including them would make the fit only better). The fit results are shown in Table~\ref{tab:bestFitsCombinedQuarkLeptonnu}, both for NO and IO. 
\begin{table}[H]%
	\centering%
	\begin{tabular}{|c||cccccccc|}%
		\hline%
		Fit & $X_a^N$ & $X_3^N$ & $\varepsilon_\phi$ & $\varepsilon_\chi$ & {min $|\lambda_{ij}^{u,d,e,\nu}|$} & {max $|\lambda_{ij}^{u,d,e,\nu}|$} & {$\chi^2$} & {$\chi^2_{{\cal O}(1)}$}\\%
		\hline
		\hline
		QL$\nu_D$-1 (NO) & 6 & 6 & 0.026 & 0.012 & $1/2.9$ & 2.6 & 0.5 & 10\\%
		QL$\nu_D$-2 (NO) & 6 & 6 & 0.024 & 0.013 & $1/2.6$ & 2.2 & 18 & 9\\%
		QL$\nu_D$-3 (NO) & 5 & 5 & 0.022 & 0.006 & $1/3.1$ & 3.8 & 1.0 & 13\\%
		QL$\nu_D$-4 (NO) & 5 & 5 & 0.021 & 0.006 & $1/2.5$ & 2.4 & 18 & 9\\%
		\hline
		{\small QL$\nu_D$ (IO) }& {\small 6} & {\small 6} & {\small 0.015} & {\small 0.013} & {\small $1/9.1$} & {\small 5.5} & {\small 18} & {\small 25}\\%
		\hline%
	\end{tabular}%
	\caption{Best fits of the combined quark and lepton sector including CKM phase and Dirac neutrinos, with normal ordering (NO) or inverted ordering (IO). The complete set of parameters can be found in Table~\ref{tab:fitResults}.}%
	\label{tab:bestFitsCombinedQuarkLeptonnu}%
\end{table}%
\noindent As expected, good fits are obtained only for equal charges $X^N_a = X^N_3 = 5 \div 6$. There is clearly a strong preference for NO, as can be seen in both quality parameters  $\chi^2$ and  $\chi^2_{{\cal O}(1)}$ (and the smallest/largest $\lambda_{ij}$). According to our quality requirement $\chi^2_{{\cal O}(1)} < 20$, we should actually discard the IO possibility, since all fits with inverted mass ordering violate this criterion, and we include it just for illustrative purposes.  

Comparing to the fit of quark and charged lepton sector only (cf. Table~\ref{tab:bestFitsQuarkAndChargedLeptonsCP}), one can see that including neutrinos makes the fits slightly worse, but still with ${\cal O}(1)$ coefficients between $1/3$ and 3. The fits determine all neutrino parameters, and we obtain predictions for the absolute mass scales and two important observables, the sum of masses $\Sigma m_i$ as probed by satellite telescopes, and the effective neutrino mass $m_\beta = \sqrt{\sum_i m_i^2 |U_{ei}|^2}$ as measured in the $\beta$-decay spectrum close to the endpoint. All predictions are summarized in Table~\ref{predictions}. 
\begin{table}[H]
	\centering
		\begin{tabular}{|c||ccccc|}
		\hline
		Fit & {$m_1$~[meV]} & {$m_2$~[meV]} & {$m_3$~[meV]} & {$\sum m_{i}$~[meV]} & {$m_\beta$~[meV]}\\
		\hline
		\hline
		QL$\nu_D$-1  & 0.5 & 8.6 & 50 & 59 & 9 \\
		QL$\nu_D$-2  & 4.6 & 9.6 & 50 & 64 & 10 \\
		QL$\nu_D$-3  & 0.4 & 8.6 & 50 & 59 & 9 \\ 
		QL$\nu_D$-4  & 0.4 & 8.6 & 50 & 59 & 9 \\
		\hline
	\end{tabular}
	\caption{\label{predictions} Predictions for neutrino masses and observables for the NO fits in Table~\ref{tab:bestFitsCombinedQuarkLeptonnu}. }
\end{table}
\noindent Since in contrast to the quark sector there are predictions for observables that are not yet measured, we also give a range for these predictions scanning over many fits with $X^N_a = X^N_3 = 5,6$ on which we only impose the (somewhat arbitrary) condition that $\chi^2 < 20$ and the quality requirement $\chi^2_{{\cal O}(1)} < 20$ (which excludes IO). In this way we obtain predictions for the ranges of $\sum m_i$ and $m_\beta$ as shown in Table \ref{predictionsWithRangesD}, where we also indicate the value preferred in most fits. 
\begin{table}[H]
	\centering
	\begin{tabular}{|c||c|c|c|}
	\hline
		Quantity & Range~[meV] & Preferred values~[meV]\\
		\hline
		\hline
		$\sum m_i$ & \numrange[range-phrase=\:--\:]{58}{110} &  \numrange[range-phrase=\:--\:]{60}{65} \\
		$m_\beta$ & \numrange[range-phrase=\:--\:]{8}{26} &  \numrange[range-phrase=\:--\:]{9}{10} \\
		\hline
	\end{tabular}
	\caption{\label{predictionsWithRangesD}Range of predictions for $\sum m_i$ and $m_\beta$ scanning over fits with Dirac Neutrino charges $X^N_a = X^N_3 = 5,6$ and $\chi^2 < 20$ and $\chi^2_{{\cal O}(1)} < 20$. In brackets indicated are the values preferred by most fits.}
\end{table}
\noindent We notice that the predicted range for $m_\beta$   is an order of magnitude below the expected future sensitivity of $m_\beta \lesssim 0.2 \, {\rm eV}$ by the KATRIN experiment~\cite{KATRIN}. The prediction for the neutrino mass sum  is consistent with present bound by PLANCK giving $\sum m_i < 0.12 \, {\rm eV}$~\cite{PLANCK} and in the reach of the EUCLID satellite that is expected to measure  $\sum m_i$ with an error of about $0.05 \, {\rm eV}$~\cite{EUCLID1, EUCLID2}. Note that the lower bound on the predicted range of $\sum m_i$ essentially saturates the minimal value that is obtained for a massless lightest neutrino, which (including $1\sigma$ errors) is given by 58 meV  for normal ordering.

\bigskip

Finally, we discuss the case of Majorana Neutrinos. In addition to the neutrino Yukawa coupling, the Lagrangian contains a Majorana mass term, ${\cal L}_\nu =  L^T Y_\nu N H + 1/2 \, N^T M_\nu N +{\rm h.c.}$ The Yukawa matrix $Y_\nu$ is the same as in Eq.~\eqref{eq:ynu}, while the Majorana mass matrix can be obtained as 
\begin{align}%
M_\nu & = M \begin{pmatrix} \kappa_{11} \eps_\phi^2 \eps_\chi^{\abs{2+2X_a^N}} & \kappa_{12} \eps_\phi^2 \eps_\chi^{\abs{2X_a^N}} & \kappa_{13} \eps_\phi \eps_\chi^{\abs{1+X_a^N+X_3^N}}\\
\kappa_{12} \eps_\phi^2 \eps_\chi^{\abs{2X_a^N}} & \kappa_{22} \eps_\phi^2 \eps_\chi^{\abs{2X_a^N-2}} & \kappa_{23} \eps_\phi \eps_\chi^{\abs{X_a^N+X_3^N-1}}\\
\kappa_{13} \eps_\phi \eps_\chi^{\abs{1+X_a^N+X_3^N}} & \kappa_{23} \eps_\phi \eps_\chi^{\abs{X_a^N+X_3^N-1}} & \kappa_{33} \eps_\chi^{\abs{2X_3^N}} 
\end{pmatrix} \, , 
\label{eq:Mheavy}
\end{align}%
where we factored out a single mass scale $M$ that is taken of the order of the usual see-saw scale, $M \sim 10^{14} \GeV$. One can therefore integrate out the heavy singlets and get light neutrino masses from the Weinberg operator $y_{ij}/M (L_i H) (L_j H)$, according to the type-I seesaw formula 
\begin{align}
m_\nu^M =  v^2 Y_\nu M_\nu^{-1} Y_\nu^T \, .  
\label{eq:seesaw}
\end{align}  
Notice that the 1-2 entry of $M_\nu$ without any $\phi$ insertion vanishes because of the necessary $SU(2)$ anti-symmetrization, and therefore picks up an additional $\eps_\phi^2$ suppression. It turns out that this extra suppression spoils the naive EFT spurion analysis of the Weinberg operator using only the charges of $L_a, L_3$ (since negative powers of $\phi$ appear in the UV theory), and one has to use Eq.~\eqref{eq:seesaw} to calculate  $m_\nu^M$. We first assume that $X^N_A \geq 1$ and $X^N_3 \geq 0$, so that one can drop the absolute values and obtain for the parametric structure of the light neutrino mass matrix
\begin{align}
m_\nu^M \sim \frac{v^2}{M} \begin{pmatrix} \eps_\chi^4/\eps_\phi^2 & \eps_\chi^2/\eps_\phi^2 & \eps_\chi^3/\eps_\phi \\%
		\eps_\chi^2/\eps_\phi^2 & 1/\eps_\phi^2 & \eps_\chi/\eps_\phi \\%
		 \ \eps_\chi^3/\eps_\phi &  \eps_\chi/\eps_\phi &  \eps_\chi^2  \end{pmatrix}   
		 \sim \begin{pmatrix}  \varepsilon^2 & 1 & \eps^2\\%
		1 & 1/\varepsilon^2 & 1\\%
		\varepsilon^2 & 1 & \varepsilon^2 \end{pmatrix} \, , %
\end{align}
where $\eps \sim \eps_\phi \sim \eps_\chi$ (notice that the charges $X^N_{a,3}$ drop out). Such a structure is clearly ruled out, since it gives singular values  $\{ \eps^2, \eps^2, 1/\eps^2 \}$, which would imply normal hierarchy along with a parametric prediction for the ratio of mass differences $\Delta m^2_{21} / \Delta m^2_{31} \sim \eps^4 \times \eps^4$ that is way too small. Moreover, one can check that also different charge assignments for $X^N_{a,3}$ do not allow to obtain a Majorana neutrino mass matrix that leads to a good fit, besides losing predictivity. Indeed the main theoretical advantage of Majorana neutrinos over Dirac neutrinos would be a scenario in which the effective Majorana mass matrix does not depend on the details of the UV physics, i.e. the choice of $X^N_{a,3}$. 

\bigskip

We conclude this section with the observation that the Majorana scenario would work perfectly if not for the vanishing of the leading 1-2 entry in the heavy mass matrix in Eq.~\eqref{eq:Mheavy}. Indeed, if this entry would be given by $\kappa_{12} \eps_\chi^{|2 X^N_a|}$, and $X^N_a \geq 1$, $X^N_3 \geq 0$, the   effective light neutrino mass matrix would be given by (the dependence on $X^N_{a,3} $ drops out again)
\begin{align}
m_\nu^M \sim \frac{v^2}{M} \begin{pmatrix} \eps_\chi^4 \eps_\phi^2 & \eps_\chi^2 & \eps_\chi^3 \eps_\phi \\%
		\eps_\chi^2  & \eps_\phi^2 & \eps_\chi \eps_\phi \\%
		 \ \eps_\chi^3 \eps_\phi &  \eps_\chi \eps_\phi &  \eps_\chi^2  \end{pmatrix}   
		 \sim \begin{pmatrix}  \varepsilon^6 & \eps^2 & \eps^4\\%
		\eps^2 & \varepsilon^2 & \eps^2 \\%
		\varepsilon^4 & \eps^2 & \varepsilon^2 \end{pmatrix} \, , %
\label{mnuMesti}		
\end{align}
which apart from the subleading $11,13,31$ entries has only very mild $\eps_\chi/\eps_\phi$ hierarchies and suggests a very good fit to neutrino observables. Note this absence of hierarchies is actually a prediction of the quark and charged lepton sector, which requires equal charges for the left-handed doublets $L_a$ and $L_3$, and order parameters of similar size, $\eps_\chi \sim \eps_\phi$. If therefore the 1-2 elements were symmetric instead of anti-symmetric, all low-energy mass matrices would follow the same hierarchical pattern, differing only in the $U(1)_F$ charge assignment of the third generation, which is 0 for $Q_3, U_3, E_3$ and 1 for $D_3, L_3$. Thus the light $2\times 2$ sub-block would be the same in all fermion sectors, and only the third coloum/row would differ by powers of $\eps_\chi$,  giving 
\begin{align}
m_{\{ u, d, e, \nu \}} \sim \begin{pmatrix} 0 &   \eps^2 & 0 \\%
		\eps^2 & \eps^2 & \{ \eps, \eps^2, \eps, \eps^2 \}  \\%
		0  & \{ \eps, \eps, \eps^2, \eps^2 \} & \{ 1, \eps, \eps, \eps^2 \} \end{pmatrix} \, , %
\end{align}
 where we neglected the mild $\eps_\chi/\eps_\phi$ hierarchy that is responsible for e.g. the Cabibbo angle. As we discuss in the following section, this simple pattern allows for an excellent fit to all fermion observables, and the necessary 1-2 symmetric structure can be obtained when considering the (discrete) dihedral group $D_6$ instead of $SU(2)$ as  flavor symmetry, which closely resembles the $SU(2)$ structure apart from a sign flip in the 1-2 entries.

\section{A $D_6 \times U(1)$ Model of Flavor}
\label{D6}
In this section we consider the same framework with a $D_6 \times U(1)$ flavor symmetry, which closely resembles the $U(2)$ case. We first introduce some $D_6 \simeq D_3 \times Z_2$ group theory and discuss the resulting flavor structure of quark and charged lepton masses, as well as the Weinberg operator. After some brief analytical considerations for the resulting predictions for neutrino observables, we perform a numerical fit to all fermion observables and conclude with  a discussion of the phenomenological implications.  

\subsection{Setup}
\setcounter{equation}{0}

As we have just discussed, we want to mimic the structure of $U(2)$ within a discrete flavor group that allows for a symmetric singlet contraction of two doublets. The simplest such group is the dihedral group $D_3$, the symmetry group of an equilateral triangle, which is discussed in detail in Appendix A. This group is actually a subgroup of $SO(3)$ and not of its double cover $SU(2)$, and it is  isomorphic to the permutation group $S_3$. It features two one-dimensional representations ${\bf 1}$ and ${\bf 1^\prime}$ and one two-dimensional representation ${\bf 2}$. The contraction of two doublets $\psi =( \psi_1,  \psi_2)$ and $\phi = (\phi_1, \phi_2)$ into the singlet ${\bf 1}$ is given by 
\begin{align}
(\psi \otimes \phi )_{\bf 1} = \psi_1 \phi_2 + \psi_2 \phi_1 \, .
\end{align}
Therefore we could simply assign the SM and spurion fields to  $D_3$  representations that follow the $SU(2)$ ones, i.e. the doublets ${\bf 10}_a, {\bf \overline{5}}_a, \phi_a$ are in a ${\bf 2}$ of $D_3$ and all other fields are total singlets. 
However, in contrast to $SU(2)$ the product of two doublets also containts a doublet, so that three doublets can be contracted to a singlet as
\begin{align}
(\psi \otimes \phi  \otimes \chi)_{\bf 1} = \psi_1 \phi_1 \chi_1 + \psi_2 \phi_2 \chi_2 \, .
\end{align}
This implies that in contrast to the $SU(2)$ model a large 1-1 entry is generated, for example in the up-sector by the operator 
\begin{align}
{\cal L} \supset \frac{1}{\Lambda^2} \left( \phi \otimes Q_a \otimes U_a \right)_{\bf 1} H \chi = \frac{1}{\Lambda^2} \left( \phi_1 Q_1  U_1 + \phi_2 Q_2  U_2 \right) H \chi = \eps_\phi \eps_\chi Q_1 U_1 H \, , 
\end{align}
which would be no longer negligible and thus would completely spoil the hierachical structure. In order to suppress this entry, we would like to mimic the $SU(2)$ structure in which such a contraction is forbidden by the $Z_2$ center of $SU(2)$, under which the doublets are odd and the singlet is even.  Therefore, we consider\footnote{Note we cannot use the double cover $\tilde{D}_3$ (which is an actual subgroup of $SU(2)$) for this purpose, since that doublet ${\bf 2}_x$ that contains no singlet in its cubic contraction, contains the singlet in its antisymmetric quadratic contraction. } $D_3 \times Z_2$ which is isomorphic to $D_6$, the symmetry group of a regular hexagon (see Appendix A for details), and finally make the charge assignment as in Table~\ref{tab:chargesD6}.  
\begin{table}[h]
\centering
\begin{tabular}{cccccccc}
\toprule
& ${\bf 10}_a$ & ${\bf \overline{5}}_a$ &${\bf 10}_3$ & ${\bf \overline{5}}_3$  &  $H$ & $\phi_a$ & $\chi$ \\
\midrule
$D_3 \times Z_2$ & ${\bf 2_-}$ & ${\bf 2_-}$  & ${\bf 1_+}$ & ${\bf 1_+}$ & ${\bf 1_+}$ & ${\bf 2_-}$ & ${\bf 1_+}$ \\    
$U(1)_F$ & $1$ & $1$ & $0$ & $1$ & $0$ & $-1$ & $-1$    \\
\bottomrule
\end{tabular}
\caption{The field content and $(D_6 \simeq D_3 \times Z_2) \times U(1)_F$ quantum numbers. \label{tab:chargesD6}}
\end{table}
The additional $Z_2$ factor ensures that the contraction of three ${\bf 2_-}$ doublets does not contain the total singlet ${\bf 1}_+$, and in the quark and charged lepton sector we obtain the very same spurion analysis as for $U(2)$ in Section 2 (see Eq.~\eqref{eq:yf}), except for the sign in the 1-2 entry:
\begin{gather}
Y_u  \approx
\begin{pmatrix}
\lambda_{11}^u \eps_\phi^2  \eps_\chi^4 & \lambda_{12}^u \eps_\chi^2  & \lambda_{13}^u \eps_\phi  \eps_\chi^2  \\ 
\lambda_{12}^u \eps_\chi^2 & \lambda_{22}^u \eps_\phi^2  & \lambda_{23}^u \eps_\phi \\
\lambda_{31}^u \eps_\phi  \eps_\chi^2 & \lambda_{32}^u \eps_\phi & \lambda_{33}^u
\end{pmatrix} \, , \qquad
Y_d  \approx
\begin{pmatrix}
\lambda_{11}^d \eps_\phi^2  \eps_\chi^4 & \lambda_{12}^d \eps_\chi^2  & \lambda_{13}^d \eps_\phi  \eps_\chi^3  \\ 
 \lambda_{12}^d \eps_\chi^2 & \lambda_{22}^d \eps_\phi^2  & \lambda_{23}^d \eps_\phi \eps_\chi \\
\lambda_{31}^d \eps_\phi  \eps_\chi^2 & \lambda_{32}^d \eps_\phi & \lambda_{33}^d \eps_\chi 
\end{pmatrix} \, , \\
Y_e  \approx
\begin{pmatrix}
\lambda_{11}^e \eps_\phi^2  \eps_\chi^4 & \lambda_{12}^e \eps_\chi^2  & \lambda_{13}^e \eps_\phi  \eps_\chi^2  \nonumber \\ 
 \lambda_{12}^e \eps_\chi^2 & \lambda_{22}^e \eps_\phi^2  & \lambda_{23}^e \eps_\phi  \\
\lambda_{31}^e \eps_\phi  \eps_\chi^3 & \lambda_{32}^e \eps_\phi \eps_\chi & \lambda_{33}^e \eps_\chi 
\end{pmatrix} \, .
\label{eq:yfD6}
\end{gather}
In the neutrino sector we work with the effective Weinberg operator $y_{ij}/M (L_i H) (L_j H)$, which can be induced by the type-I seesaw mechanism as discussed in the previous section. Its parametric structure is predicted in terms of the $D_6 \times U(1)_F$ quantum numbers of the charged leptons, which gives for the light Majorana neutrino mass matrix 
\begin{align}%
	m_{\nu} \approx \frac{v^2}{M} \begin{pmatrix} \lambda_{11}^\nu \varepsilon_\chi^4 \eps_\phi^2 & \lambda_{12}^\nu \varepsilon_\chi^2 & \lambda_{13}^\nu \eps_\phi \varepsilon_\chi^3  \\%
						   \lambda_{12}^\nu \varepsilon_\chi^2 & \lambda_{22}^\nu \varepsilon_\phi^2 & \lambda_{23}^\nu \varepsilon_\phi \varepsilon_\chi\\%
						   \lambda_{13}^\nu \eps_\phi \varepsilon_\chi^3  & \lambda_{23}^\nu\varepsilon_\phi \varepsilon_\chi & \lambda_{33}^\nu \varepsilon_\chi^2 \end{pmatrix} \, .
\end{align}%
Here we have used the same vacuum expectation values as before
\begin{align}
\langle \phi \rangle & =  \begin{pmatrix} \eps_\phi \Lambda \\ 0 \end{pmatrix} \, , &
 \langle \chi \rangle & = \eps_\chi \Lambda \, ,
\end{align}
although in contrast to the $SU(2)_F$ case we cannot use $D_6$ transformations in order to assume this VEV for $\phi$ without loss of generality. Therefore, we provide an explicit scalar potential in Appendix~\ref{Vscal} with only one additional scalar field that generates dynamically the above VEVs~\footnote{Also a tiny VEV along the lower component of $\phi$ is generated, which however is small enough to give only negligible contributions to masses and mixings.}. Altogether, we obtain to good approximation the mass matrices
\begin{align}
m_u  & \approx
v \begin{pmatrix}
0 & \lambda_{12}^u \eps_\chi^2  &0   \\ 
\lambda_{12}^u \eps_\chi^2 & \lambda_{22}^u \eps_\phi^2  & \lambda_{23}^u \eps_\phi \\
0 & \lambda_{32}^u \eps_\phi & \lambda_{33}^u
\end{pmatrix} \, , & 
m_d  & \approx
v \begin{pmatrix}
0  & \lambda_{12}^d \eps_\chi^2  & 0 \\ 
 \lambda_{12}^d \eps_\chi^2 & \lambda_{22}^d \eps_\phi^2  & \lambda_{23}^d \eps_\phi \eps_\chi \\
0 & \lambda_{32}^d \eps_\phi & \lambda_{33}^d \eps_\chi 
\end{pmatrix} \, , \\
m_e  & \approx
v \begin{pmatrix}
0 & \lambda_{12}^e \eps_\chi^2  &0  \nonumber \\ 
 \lambda_{12}^e \eps_\chi^2 & \lambda_{22}^e \eps_\phi^2  & \lambda_{23}^e \eps_\phi  \\
0 & \lambda_{32}^e \eps_\phi \eps_\chi & \lambda_{33}^e \eps_\chi 
\end{pmatrix} \, , & 
m_{\nu} & \approx \frac{v^2}{M} \begin{pmatrix}0 & \lambda_{12}^\nu \varepsilon_\chi^2 &0  \\%
						   \lambda_{12}^\nu \varepsilon_\chi^2 & \lambda_{22}^\nu \varepsilon_\phi^2 & \lambda_{23}^\nu \varepsilon_\phi \varepsilon_\chi\\%
						  0 & \lambda_{23}^\nu\varepsilon_\phi \varepsilon_\chi & \lambda_{33}^\nu \varepsilon_\chi^2 \end{pmatrix} \, .
\label{eq:yallD6}
\end{align}
As discussed in the previous section, this model has the remarkable feature that the hierarchies in the quark and charged lepton sector require $\eps_\phi \sim \eps_\chi$, and therefore naturally gives rise to an approximately anarchic neutrino mass matrix with generically large mixing angles.

Before we perform a numerical fit, we proceed with some analytical considerations. In the quark and charged lepton sector the analysis of the previous section is unaltered, since the flipped sign in the 1-2 entry does not play a role at leading order. In the neutrino sector we have 4 real parameters, which will enter the PMNS matrix together with three charged lepton rotations angles controlled by a single free real parameter $s_{23}^{Le}$, see Eq.~\eqref{eq:approxrotangles}. These parameters correspond to 5 observables (3 PMNS angles + 2 squared mass differences), so up to phases all parameters are fixed and one can predict the absolute neutrino mass scales and related observables. There are 4 phases in the neutrino sector and 2 phases in the left-handed charged lepton rotations, which combine to 3 physical phases, one Dirac and two Majorana phases. To study the prediction of the overall neutrino mass scale, we parametrize the neutrino mixing matrix $V_\nu$ (defined by  $V_\nu^T m_\nu V = m_\nu^{\rm diag}$) in the standard CKM form multiplied with a phase matrix $P_\nu = {\rm diag} (e^{i \alpha_1}, e^{i \alpha_2}, 1) $ from the right and a phase matrix $P^\prime$ from the left. Inverting the defining equation, we get from the vanishing 11 and 13 entries the two equations
\begin{align}
c_{12, \nu}^2 \frac{m_1}{m_3} e^{- 2 i (\alpha_1 + \delta_\nu)} + s_{12, \nu}^2  \frac{m_2}{m_3} e^{- 2 i (\alpha_2 + \delta_\nu)}  + \frac{s_{13,\nu}^2}{c_{13,\nu}^2} = 0 \, , \\
 \frac{m_1}{m_3} e^{-  i (2 \alpha_1 + \delta_\nu)} - \frac{m_2}{m_3} e^{-  i (2 \alpha_2 + \delta_\nu)}  + \frac{s_{13,\nu} c_{23,\nu} }{c_{13,\nu}^2 c_{12,\nu} s_{12,\nu} s_{23,\nu}} = 0 \, . 
\end{align}
This leads to the inequalities
\begin{gather}
\label{eqA}
\left| 1 - \frac{c_{12, \nu}^2 }{s_{12, \nu}^2} \frac{m_1}{m_2} \right| \le \frac{s_{13, \nu}^2}{s_{12,\nu}^2 c_{13, \nu}^2} \frac{m_3}{m_2} \le  1 + \frac{c_{12, \nu}^2 }{s_{12, \nu}^2} \frac{m_1}{m_2} \, , \\
 1 - \frac{m_1}{m_2}  \le \frac{s_{13,\nu} c_{23,\nu} }{c_{13,\nu}^2 c_{12,\nu} s_{12,\nu} s_{23,\nu}} \frac{m_3}{m_2} \le  1 + \frac{m_1}{m_2}   \, .
\end{gather}
The angles in the neutrino sector $s_{ij,\nu}$ are connected to the observed PMNS mixing angles through $V_{\rm PMNS} = (V^e_L)^T  V_\nu$. Since the 1-2 rotation in the charged lepton sector is small, $\sim \sqrt{m_e/m_\mu} \approx 0.07$, we have to good approximation $s_{12, \nu} \approx s_{12}$, but 2-3 rotations in the charged lepton sector are large, so that both $\theta_{23}$ and $\theta_{13}$ generically receive large contributions from the charged lepton sector. Nevertheless one can easily verify that Eq.~\eqref{eqA} cannot be satisfied for inverted mass ordering, while for normal ordering one can obtain an upper bound on the lightest neutrino mass $m_1$, by maximizing the neutrino mixing angles $s_{12,\nu}$ and $s_{13,\nu}$ with a suitable choice of phases. If one neglects the charged lepton contribution to $s_{12}$, one can show that $m_1  \le  11 \, {\rm meV}$, which in turn leads to upper bounds $\sum m_i \le 76 \, {\rm meV}$, $m_\beta \le 14 \, {\rm meV}$ and $m_{\beta \beta} \le 13 \, {\rm meV}$. This estimate is confirmed by the numerical analysis in the next section.

\subsection{Numerical Fit}
\label{D6fit}
We now perform a simultaneous fit to quark, charged lepton and neutrino sector including a phase in $\lambda^u_{33}$ as in the last section (for simplicity we omit phases in the neutrino sector, including them would make the fit only better). The fit results are shown in Table~\ref{tab:bestFitsCombinedQuarkLeptonnuM}, and include also the effective suppression scale $M$ of Weinberg operator, which is of the order of $10^{11} \GeV$. The fit is even better compared to Dirac Neutrinos (cf. Table~\ref{tab:bestFitsCombinedQuarkLeptonnu}), with all ${\cal O}(1)$ parameters roughly between 0.4 and 2.
\begin{table}[H]
	\centering
		\begin{tabular}{|c||cccccc|c|}
		\hline
		Fit & {$\varepsilon_\phi$} & {$\varepsilon_\chi$} & {min $|\lambda_{ij}^\mathrm{u,d,\ell}|$} & {max $|\lambda_{ij}^\mathrm{u,d,\ell}|$} & {$\chi^2$} & {$\chi^2_{{\cal O}(1)}$} & $M$ [$10^{11} \GeV$]\\
		\hline
		\hline
		QL$\nu_M$-1 & 0.025 & 0.009 & $1/2.8$ & 2.1 & 0.7 & 7.9 & 4.1 \\
		QL$\nu_M$-2 & 0.024 & 0.009 & $1/2.6$ & 1.9 & 18 & 6.3 & 3.3 \\
		\hline
	\end{tabular}
	\caption{Best fits for the $D_6 \times U(1)$ model  including CKM phase and Majorana neutrinos. The complete set of parameters can be found in Table~\ref{tab:fitResults}. \label{tab:bestFitsCombinedQuarkLeptonnuM}}
\end{table}

The corresponding predictions for the neutrino masses $m_i$, its sum $\sum m_i$, the neutrino mass $m_\beta$ and the ``effective Majorana mass" $m_{\beta \beta} = \left| \sum U_{ei}^2 \, m_i \right|$ measured in neutrinoless double-beta decay  are shown in Table~\ref{tab:predictionsD6MajoranaModel}. As expected from the analytical considerations, only a normal hierarchy for the neutrino masses is viable. The predicted values for $\sum m_i$ and $m_\beta$ are similar to the ones in the Dirac Neutrino case (cf.~Table~\ref{predictions}), while the effective Majorana mass is well below the expected sensitivities even in near future neutrinoless double-beta decay experiments~\cite{DellOro:2016tmg}.
\begin{table}[H]
	\centering
	\begin{tabular}{|c||cccccc|}
		\hline
		Fit & {$m_1$~[meV]} & {$m_2$~[meV]} & {$m_3$~[meV]} & {$\sum m_i$~[meV]} & {$m_\beta$~[meV]} & {$m_{\beta \beta}^\mathrm{max}$~[meV]}\\
		\hline
		\hline
		QL$\nu_M$-1 & \num{1.0} & \num{8.7} & \num{50} & \num{60} & $8.8$ & $4.4$\\
		QL$\nu_M$-2 & \num{1.5} & \num{8.8} & \num{50} & \num{60} & $8.9$ & $4.9$\\
		\hline
	\end{tabular}
	\caption{Predictions for neutrino masses and observables for the fits in Table~\ref{tab:bestFitsCombinedQuarkLeptonnuM}. Since the  prediction for $m_{\beta \beta}$ strongly depends on possible  phases in the PMNS matrix, here we display the maximal possible value $m_{\beta \beta}^\mathrm{max}$.  \label{tab:predictionsD6MajoranaModel}}

\end{table}
Finally, we also give a range for the observables scanning over many fits  on which we only impose that $\chi^2 < 20$ and $\chi^2_{{\cal O}(1)} < 20$. In this way we obtain predictions for $\sum m_i$, $m_\beta$ and $m_{\beta \beta}$ lying in the ranges shown in Table~\ref{rangeD6}, where we also indicate the value preferred in most fits. This result agrees well with our estimate in the last section, where we have also included phases, so we expect the upper bounds on the mass scales to be approximately valid even when including phases in the numerical fit (the lower bounds again saturate the limit obtained from taking the lightest neutrino massless).
\begin{table}[H]
	\centering
	\begin{tabular}{|c||c|c|c|}
	\hline
		Quantity & Range~[meV] & Preferred values~[meV]\\
		\hline
		\hline
		$\sum m_i$ & \numrange[range-phrase=\:--\:]{59}{78} & $60$, $70$\\
		$m_\beta$ & \numrange[range-phrase=\:--\:]{8}{15} &  \numrange[range-phrase=\:--\:]{9}{10},  \numrange[range-phrase=\:--\:]{11}{12}\\
		$m_{\beta \beta}^\mathrm{max}$ & \numrange[range-phrase=\:--\:]{3}{16} & $5$, $9$\\
		\hline
	\end{tabular}
	\caption{Range of predictions for $\sum m_i$, $m_\beta$ and $m_{\beta \beta}$ scanning over fits with $\chi^2 < 20$ and $\chi^2_{{\cal O}(1)} < 20$. The last column indicates the values preferred by most fits.  \label{rangeD6}}
\end{table}

\bigskip

We conclude this section with a discussion of the phenomenological implications of our model. As we have seen, the flavor sector itself gives rise to quite narrow predictions for observables in the neutrino sector, which are however far below the present experimental sensitivities. In order to obtain other experimental signals, we have to rely on new low-energy dynamics besides the SM. The natural candidate for such new degrees of freedom are the fields at the cut-off scale $\Lambda$,  which we have not specified so far (in particular the radial components of the flavons $\phi$ and $\chi$ naturally get a mass at that scale). However, effects of these fields and other dynamics related to the UV completion are suppressed by powers of $1/\Lambda$, and there is no reason that $\Lambda$ is sufficiently close to the electroweak scale in order to give rise to sizable deviations from the SM. Still, it would be interesting to consider an explicit UV completion of the present model to study the structure of these effects in detail.

Another option for light dynamics, which is essentially model-independent and  well-motivated, is provided by the pseudo-scalars in the flavon fields. If there is no explicit breaking of the $U(2)_F$ symmetry, the associated Goldstone bosons are exactly massless, apart from a linear combination that can be identified with the QCD axion, which solves the strong CP problem and gets a mass from non-perturbative effects. The easiest way to get rid of the orthogonal massless Goldstones is replacing $SU(2)_F$ by a discrete subgroup, which is another advantage of the $D_6 \times U(1)$ model discussed in this section. In this case there a single Goldstone boson associated with the $U(1)_F$ factor that can naturally serve as the QCD axion, as we are going to discuss in the next section.

\section{The $U(2)$ Axiflavon }
\setcounter{equation}{0}

As originally proposed in Ref.~\cite{Wilczek}, a Goldstone boson arising from the breaking of global flavor symmetries could play the role of the QCD axion. Indeed any Goldstone of a $U(1)$ symmetry with a QCD anomaly will solve the strong CP problem, and one can demonstrate (see Ref.~\cite{Axiflavon}) that there is a non-zero $SU(3)_c \times SU(3)_c \times U(1)_F$ anomaly in any flavor  model where the determinants of up-down and down-quark mass matrices are controlled dominantly by the $U(1)_F$ symmetry factor. In the present model this is indeed the case as $\det m_u \sim \eps_\chi^4$ and  $\det m_d \sim \eps_\chi^5$, due to the presence of the approximate texture zeros, see Eq.~\eqref{eq:yallD6}. Moreover, if also the determinant of the charged lepton mass matrix depends only on the $U(1)_F$ breaking, the ratio of electromagnetic and color anomaly coefficients $E/N$ is expected to be a rational number close to $8/3$~\cite{Axiflavon}. In the present model the $U(1)_F$ charge assignment is actually compatible with $SU(5)$,  so it is clear that we get exactly $E/N = 8/3$, as in minimal DFSZ~\cite{DFSZ1, DFSZ2} and KSVZ models~\cite{KSVZ1, KSVZ2}, and thus the same axion couplings to photons. 

In this section we will calculate the axion couplings to photons and fermions, concentrating on the flavor-violating couplings to fermions, which follow from the hierarchical structure of fermion masses and mixings. In particular, axion couplings to nucleons and electrons are fixed in terms of the $U(1)_F$ charges,  while flavor-violating couplings to quarks and leptons are controlled by the unitary rotations that diagonalize the Yukawa matrices. Their parametric suppression is determined by the $U(2)_F$ quantum numbers, and their numerical value by the fit to fermion masses and mixings.  We then study the phenomenology of this axion, finding that the strongest constraints on the axion mass (or equivalently the $U(1)_F$ breaking scale) come from astrophysical constraints (as in the minimal DFSZ and KSVZ models), since flavor-violating axion couplings to light quarks are strongly suppressed by the approximate $SU(2)_F$ structure.
\subsection{Axion Couplings}
We begin by identifying the axiflavon as the Goldstone boson arising from the spontaneous breaking of $U(1)_F$ induced by the VEVs of $\phi$ and $\chi$. In general, the Goldstone is a linear combination of the phases $a_i$ of the scalar fields $\phi_i$ with charge $X_i$ and (real) VEV $V_i$, given by
\begin{align}
a & = \sum_i \frac{X_i V_i a_i}{\sqrt{\sum X_j^2 V_j^2}} \, .
\end{align}
Thus, we find that $\chi$ and $\phi$ contain the Goldstone as (we ignore the radial mode)
\begin{align}
\chi & = \eps_\chi \Lambda e^{- ia(x)/\sqrt2 V}  \, , & \phi & = \begin{pmatrix} \eps_\phi \Lambda \\ 0 \end{pmatrix} e^{- ia(x)/\sqrt2 V} \, ,
\end{align}
where we have defined the $U(1)_F$ breaking scale $V\equiv \sqrt{\eps_\chi^2 + \eps_\phi^2} \, \Lambda$.

The couplings of $a$ to fermions can be obtained by inserting the above expressions for $\chi$ and $\phi$ into the effective Yukawa Lagrangian given by Eq~\eqref{Lu} for the up sector and the analogous terms in the down- and charged lepton sector.  It is then convenient to change field basis by performing a $U(1)_F$ transformation of the fermion fields 
\begin{align}
f \to f e^{i X_f a(x)/\sqrt2 V } \, , 
\end{align} 
which will remove the $a(x)$ dependence from the Yukawa sector, because of $U(1)_F$ invariance. Since this transformation is anomalous, it will generate axion couplings to gauge field strengths, and since it is local it will modify fermion kinetic terms. The resulting couplings to gluon and photon fields strengths are given by 
\begin{align}
{\cal L}_{\rm anom} = N \frac{a(x)}{\sqrt2 V} \frac{\alpha_s}{4 \pi} G_{\mu \nu} \tilde{G}^{\mu \nu} + E \frac{a(x)}{\sqrt2 V} \frac{\alpha_{\rm em}}{4 \pi} F_{\mu \nu} \tilde{F}^{\mu \nu} \, , 
\end{align}
with the dual field strength $\tilde{F}_{\mu \nu} = \frac{1}{2} \eps_{\mu \nu \rho \sigma} F^{\rho \sigma}$ and the anomaly coefficients
\begin{align}
N & = \frac{1}{2} \left( 4 X_{10_a} + 2 X_{10_3} + 2 X_{10_a} + X_{10_3}  + 2 X_{{\overline{5}}_a}  + X_{{\overline{5}}_3} \right) = 9/2 \, , \\  E & = \frac{5}{3} \left(   2 X_{10_a} +  X_{10_3} \right) + \frac{4}{3}  \left( 2 X_{10_a} +  X_{10_3} \right) +  \frac{1}{3} \left(  2 X_{{\overline{5}}_a}  + X_{{\overline{5}}_3} \right) \nonumber \\
& + \left(  2 X_{{\overline{5}}_a}  + X_{{\overline{5}}_3}  \right) +  \left(  2 X_{10_a} +  X_{10_3}  \right)  = 12 \, .
 \end{align}
Thus, we obtain $E/N = 8/3$ exactly, which is just a consequence of the fact that the $U(1)_F$ charge assignment is compatible with $SU(5)$. The modification of fermion kinetic terms  leads to  axion-fermion couplings in the flavor interaction basis 
\begin{align}
{\cal L}_a & =  - \frac{\partial_\mu a}{\sqrt2 V} \sum_f  f^\dagger_i \overline{\sigma}^\mu X_{f_i} f_i   \, . 
\end{align}
In the mass basis, defined as $m_f = V_{fL} m_f^{\rm diag} (V_{fR})^\dagger$ we have
\begin{align}
{\cal L}_a & =  -  \frac{\partial_\mu a}{\sqrt2 V} \sum_{f = u,d,e}    \left[ g^L_{f_i f_j}  f^\dagger_i \overline{\sigma}^\mu f_j +   g^R_{f_i f_j} f^{c \dagger}_i \overline{\sigma}^\mu  f_i^c \right]  \, ,
\end{align}
with 
\begin{align}
g^L_{f_i f_j} & = (V_{fL})_{ki}
 X_{f_k} (V_{fL})^*_{kj}  =  X_{f_a} \delta_{ij} + (X_{f_3} - X_{f_a} )  (V_{fL})_{3i} (V_{fL})^*_{3j}  \, , \\
 g^R_{f_i f_j} & = (V_{fR})^*_{ki}
 X_{f^c_k} (V_{fR})_{kj} =  X_{f^c_a} \delta_{ij} +  (X_{f^c_3} - X_{f^c_a} )  (V_{fR})^*_{3i} (V_{fR})_{3j} \, .
\end{align}
Finally we switch to Dirac spinor notation for the fermions and introduce $f_a \equiv V/( \sqrt2 N)$ to match to the standard normalization for the anomalous couplings. These are given by
\begin{align}
{\cal L}_{\rm anom} =  \frac{a(x)}{f_a} \frac{\alpha_s}{8 \pi} G_{\mu \nu} \tilde{G}^{\mu \nu} + \frac{E}{N} \frac{a(x)}{f_a} \frac{\alpha_{\rm em}}{8 \pi} F_{\mu \nu} \tilde{F}^{\mu \nu} \, , 
\end{align}
with $E/N = 8/3$ in this model (and domain wall number $N_{\rm DW} = 2 N = 9$). The couplings to fermions are given by
\begin{align}
{\cal L}_a & =   \frac{\partial_\mu a}{2 f_a} \overline{f}_i \gamma^\mu \left[ C^V_{f_i f_j} + C^A_{f_i f_j} \gamma_5 \right] f_j \, , 
\label{acouplings}
\end{align}
with 
\begin{align}
C^V_{f_i f_j}  & = \frac{- g^L_{f_i f_j} + g_{f_j f_i}^R}{2 N} =  \frac{X_{f^c_a} - X_{f_a}}{2N} \delta_{ij} + \frac{X_{f^c_3} - X_{f^c_a}}{2N} \eps^{f}_{R,ij} - \frac{ X_{f_3} - X_{f_a} }{2N}  \eps^{f}_{L,ij}   \, , \\
C^A_{f_i f_j}  & = \frac{g^L_{f_i f_j} + g_{f_j f_i}^R}{2 N} =  \frac{X_{f^c_a} + X_{f_a}}{2N} \delta_{ij} + \frac{X_{f^c_3} - X_{f^c_a}}{2N} \eps^{f}_{R,ij} + \frac{ X_{f_3} - X_{f_a} }{2N}  \eps^{f}_{L,ij} \, ,
\end{align}
and the shorthand notation
\begin{align}
\eps^{f}_{L,ij} & \equiv  (V_{L}^f)_{3i} (V_{L}^f)^*_{3j} \, , & 
\eps^{f}_{R,ij} & \equiv  (V_{R}^f)_{3i} (V_{R}^f)^*_{3j} \, .
\end{align}
Note that the diagonal elements of these parameters satisfy
\begin{align}
0 & \le \eps^f_{L/R,ii} \le 1 \, , & \sum_i  \eps^f_{L/R,ii} & = 1 \, .
\end{align}
While the above expressions are valid for any axion model with PQ charges that are universal for two fermion generations\footnote{See Ref.~\cite{astrophobic} for a recent example where this structure is realized within a generalized DFSZ model, and can be used to suppress the axion couplings to nucleons and electrons.}, in the present model these expressions simplify to
\begin{align}
C^V_{u_i u_j}  & =   \frac{\eps^{u}_{L,ij} - \eps^{u}_{R,ij}}{9}  \, , &
C^A_{u_i u_j}  & = \frac{2 \delta_{ij} -  \eps^{u}_{L,ij}  -  \eps^{u}_{R,ij} }{9}  \, , \\
C^V_{d_i d_j}  & =   \frac{ \eps^{d}_{L,ij}  }{9}   \, , &C^A_{d_i d_j}  & =  \frac{2 \delta_{ij} -  \eps^{d}_{L,ij} }{9}   \, , \\
C^V_{e_i e_j}  & =   - \frac{ \eps^{e}_{R,ij}}{9}    \, , &
C^A_{e_i e_j}  & = \frac{2 \delta_{ij} -  \eps^{e}_{R,ij}}{9}   \, .
\end{align}
Using the approximate expressions in Eq.~\eqref{eq:approxrotangles}, the rotations have  the parametric structure
\begin{align}
V^u_L & \sim V^u_R \sim \begin{pmatrix}
1  & \lambda   & \lambda^7  \\ 
 \lambda & 1 & \lambda^2  \\
\lambda^3 & \lambda^2 &1
\end{pmatrix} \, , &  
V^d_L &  \sim V^e_R \sim \begin{pmatrix}
1  & \lambda   & \lambda^3  \\ 
 \lambda & 1 & \lambda^2  \\
\lambda^3 & \lambda^2 &1
\end{pmatrix} \, , &  
V^d_R &   \sim V^e_L \sim \begin{pmatrix}
1  & \lambda   & \lambda^5  \\ 
 \lambda & 1 & 1  \\
\lambda & 1 &1
\end{pmatrix} \, ,
\end{align}
so that all relevant $V_{3i}$ are CKM-like, and we have 
\begin{align}
\eps^u_{L} \sim \eps^u_{R} \sim \eps^d_{L}  \sim \eps^e_{R}  \sim \begin{pmatrix}
\lambda^6  & \lambda^5   & \lambda^3  \\ 
 \lambda^5 & \lambda^4 & \lambda^2  \\
\lambda^3 & \lambda^2 &1
\end{pmatrix} \, .
\end{align}
Therefore, the diagonal axial couplings are to very good approxmation independent of the rotations, and we get, denoting $C_{f_i} \equiv C^A_{f_i f_i}$, 
\begin{align}
C_u & = C_d = C_e = C_c = C_s = C_\mu  = \frac{2}{9} \, , & C_t & = 0 \, , & C_b & = C_\tau = \frac{1}{9} \, .
\end{align} 
The flavor-violating axion couplings are controlled by $\eps^f_{ij}$, whose numerical values, beyond the parametric suppression given above, are  known for a given fit to masses and mixings. Besides there is an overall suppression factor $1/f_a$ that is proportional to the axion mass $m_a$, with the usual conversion factor for QCD axions as obtained from Chiral Perturbation Theory~\cite{villadoro} and Lattice QCD~\cite{lattice}
\begin{align}
m_a = 5.7 \, \mu {\rm eV} \left( \frac{10^{12} \GeV}{f_a} \right) \, .
\end{align}

\subsection{Axion Phenomenology}
\label{axionpheno}
The most important constraints on fermion couplings of invisible (stable) axions (cf. Eq.~\ref{acouplings}) are summarized as an upper bound on the quantity $(m_a/{\rm coupling})$ in the first column of Table~\ref{bounds}.  These include flavor-violating $b-s$ transitions as tested in $B \to K a$ decays~\cite{BKa},  flavor-violating $s-d$ transitions contributing to $K \to \pi a$ decays~\cite{Kpia},  lepton flavor-violating $\mu-e$ transitions contributing to $\mu \to e a$~\cite{muea} and $\mu \to e a \gamma$ decays~\cite{mueaga1, mueaga2}, (flavor-diagonal) axion-electron couplings bounded by the measurement of the WD luminosity function~\cite{WDbound}, and effective axion couplings to nucleons constrained from the burst duration of the SN 1987A neutrino signal~\cite{SNboundPDG}. We did not include bounds from e.g. flavor-violating tau decays~\cite{tauea}, since they give much weaker constraints.
\begin{table}[H]
	\centering
	\begin{tabular}{|c||c|cc|c|}
	\hline
		Coupling & $m_a^{\rm max}/C$~[eV] & $m_a^{\rm max, U(2)}$~[eV] & $f_a^{\rm min, U(2)}$~[GeV]  & Constraint \\
		\hline
		\hline
		$C_{\mu e}$ & $2.1 \cdot 10^{-3} $ & 78  & $7.3 \cdot 10^4 $ & $\mu \to e a$~\cite{muea} \\
	 	$C_{b s}^V$ & $9.1 \cdot 10^{-2}$ &16  
		 & $3.6 
		  \cdot 10^{5}$ &  $B^+ \to K^+ a$~\cite{BKa} \\
		$C_{s d}^V$ & $1.7   \cdot 10^{-5}$ & 0.58  
		& $9.8
		 \cdot 10^{6}$ &  $K^+ \to \pi^+ a$~\cite{Kpia} \\
		$C_{ee}^A$ & $3.1 \cdot 10^{-3}$ & 0.014 & $4.1 \cdot 10^{8}$ & WD Cooling~\cite{WDbound} \\
		$C_{N}$ & $3.5 \cdot 10^{-3}$ & 0.0092 & $6.2 \cdot 10^{8}$ & SN1987A~\cite{SNboundPDG} \\
		\hline
	\end{tabular}
	\caption{ \label{bounds} Bounds on selected axion-fermion couplings; here $C_{\mu e} \equiv \sqrt{(C_{\mu e}^{V})^2 + (C_{\mu e}^{A})^2}$ and $C_N \equiv \sqrt{C_p^2 + C_n^2}$ denotes the effective couplings to nucleons, with axion couplings to protons and neutrons $C_{p,n}$ defined analogously to the axial vector couplings in Eq.~\eqref{acouplings}. The second column denotes the model-independent upper bounds on the ratio of $m_a/C$, where $C$ denotes the respective coupling, while the third and fourth columns contain the upper (lower) bound on $m_a$ ($f_a$) in our model, using the numerical results for the couplings of Section~\ref{D6fit}, where for explicitness we took the fit QL$\nu_M$-1 (other fits give similar constraints).}
\end{table}
We have further used the predictions of the axion couplings in our model to obtain an upper bound on $m_a$, or equivalently a lower bound on $f_a$, which is shown in Table~\ref{bounds} for the fit QL$\nu_M$-1 of the complete $D_6 \times U(1)$ model in Table~\ref{tab:bestFitsCombinedQuarkLeptonnuM} (the result for the other fits are very similar). As a result of the strong CKM protection of $s-d$ transitions $C^V_{sd} \sim \lambda^5$, the main constraint on the model comes from astrophysics, similar to flavor-universal axion models. Since the bound from WD cooling and SN1987A are comparable, and the precise value of the latter is debated in the literature (see e.g. the recent discussion in Ref.~\cite{SNboundEssig} which finds a constraint on $m_a/C$ roughly a factor 5 weaker than the PDG bound), we only take the constraint from WD cooling,  giving a upper bound on the axion mass $m_a < 14  \, {\rm meV}$. This translates into a lower bound on the cutoff $\Lambda >  1.9 \cdot 10^{10} \GeV$. The predictions for the branching ratio of $K^+ \to \pi^+ a$ decays are given
\begin{align}
{\rm BR} (K^+ \to \pi^+ a) & = 4.3 \cdot 10^{-14} \left(  \frac{m_a}{14 \, {\rm meV}} \right)^2 \, ,   
\end{align}
which is far below the future sensitivity of NA62~\cite{NA621,NA622} given the constraint from WD cooling. This is in sharp constrast to the U(1) Axiflavon  proposed in Ref.~\cite{Axiflavon} (see also Ref.~\cite{Japanese}), where the $d-s$ transition is only Cabibbo-suppressed, $C^V_{sd} \sim \lambda$, so that $K^+ \to \pi^+ a$ provides the strongest constraint on the axion mass. 

\begin{figure}[H]
\centering
\includegraphics[scale=0.33]{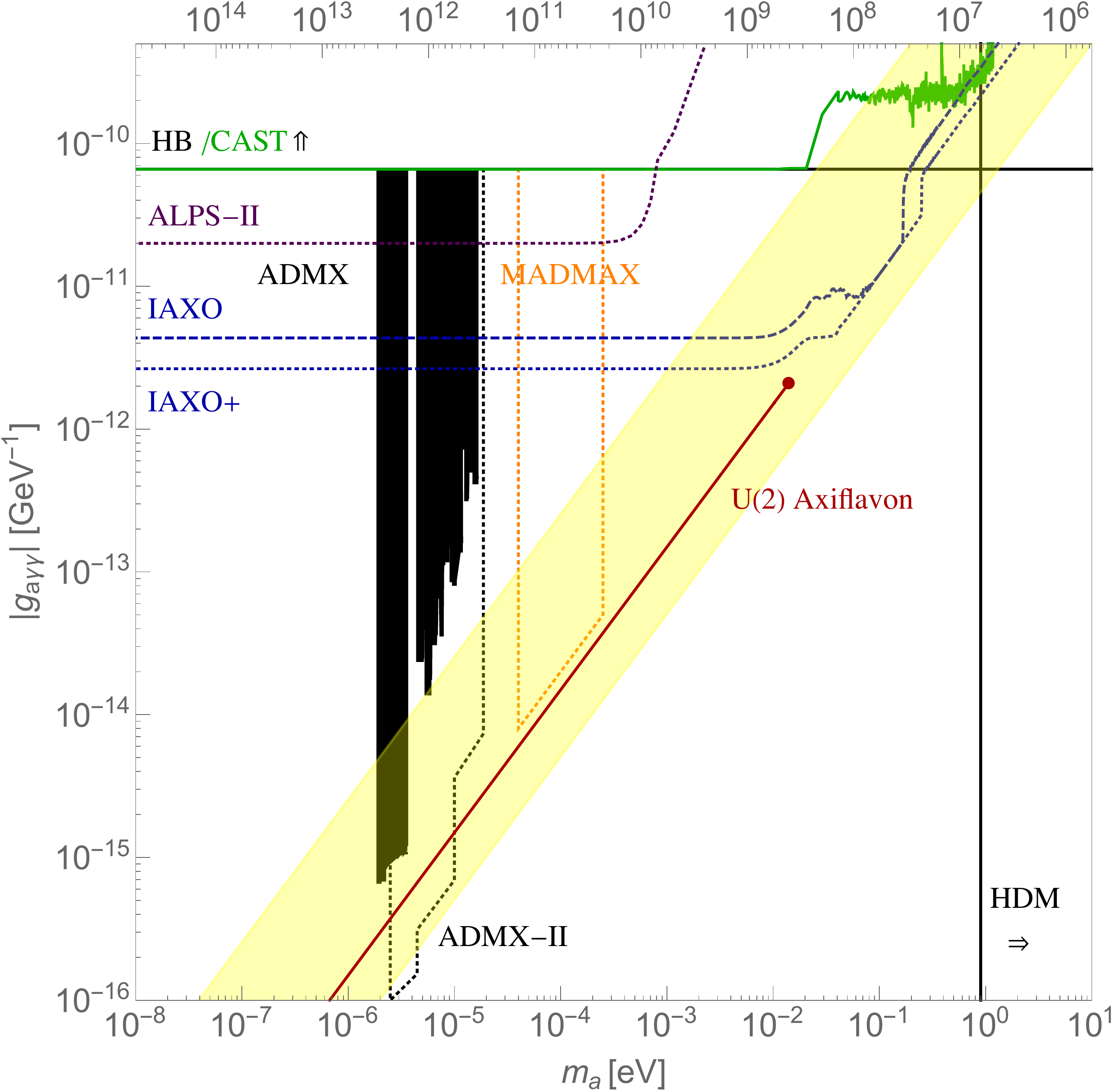}
\caption{\label{plot} 
Prediction of the axion-photon coupling as a function of the axion mass $m_a$. The yellow band denotes the usual axion band of KSVZ models with a single pair of vector-like fermions, taken from Ref.~\cite{Luca1}. The red line denotes the parameter space of the U(2) Axiflavon model, which extend up to the {\large $\bullet$} mark, denoting the bound from WD cooling, see Table~\ref{bounds}. Also shown are the bounds from structure formation excluding hot DM (HDM)~\cite{HDMbound0, HDMbound1, HDMbound2}, the bound from the evolution of Horizontal Branch (HB) stars in globular clusters~\cite{HBbounds}, the expected sensitivity of the ALPS-II experiment~\cite{ALPS2}, the present and future bounds from Axion helioscopes provided by CAST~\cite{CAST} and IAXO~\cite{IAXO1, IAXO2}, and from Axion Haloscopes like ADMX~\cite{ADMX1, ADMX2}, MADMAX~\cite{MADMAX} and the planned ADMX upgrade~\cite{ADMXfuture}.  }
 \end{figure}

The upper bound on  $m_a < 14  \, {\rm meV}$  implies that the axion is stable on cosmological scales. It is a remarkable feature of the QCD axion that it can also explain the observed Dark Matter (DM) abundance. One of the simplest scenarios is the misalignment mechanism~\cite{AxionDMmisa1, AxionDMmisa2, AxionDMmisa3}, valid when $U(1)_F$ is broken before inflation\footnote{Also cosmological scenarios with post-inflationary $U(1)_F$ breaking are viable, provided the presence of a suitable explicit breaking term to solve the domain wall problem arising from $N_{\rm DW} = 9$. This is in contrast to the U(1) Axiflavon in Ref.~\cite{Axiflavon}, where the upper bound on the axion mass from $K \to \pi a$ prevents to obtain the right amount of axion dark matter if $U(1)_F$ is broken after inflation.}. At this stage the axion is essentially massless and takes a generic field value misaligned from the vacuum value by an angle $\theta$. Around the QCD phase transition the axion potential is generated, and the axion begins to oscillate around the minimum. The energy density stored in these oscillations
can be approximately related to the present DM abundance as~\cite{PDG2016}
\begin{align}
\Omega_{\rm DM} h^2 \approx 0.12 \left( \frac{6 \, \mu {\rm eV} }{m_a} \right)^{1.165} \theta^2 \,,  
\end{align}
where $\theta \in [ - \pi, \pi ]$ is the initial misalignment angle. Thus for not too small values $\theta \gtrsim 0.1 \pi$, the natural window for axion DM is given by axion masses roughly between $(1 \div 40) \, \mu {\rm eV}$, which correspond to  axion decay constants  $f_a \sim (10^{11} \div 10^{13}) \GeV$ and a cutoff in the range\footnote{Repeating the numerical fit as in Section~\ref{D6fit} with SM input values at $10^{14}$ GeV, the $\chi^2$ and $\chi^2_{
\cal O}(1)$ get slightly worse (0.4/11 and 18/9.1 compared to 0.7/7.9 and 18/6.3 at 10 TeV, see Table~\ref{tab:bestFitsCombinedQuarkLeptonnuM}), while the overall predictions  change only marginally.} $\Lambda \sim (10^{13} \div 10^{15}) \GeV$. This range of axion masses preferred by DM through the misalignment mechanism will be probed by the ADMX upgrade in the near future~\cite{ADMXfuture}. Indeed the discovery prospects of the $U(2)$ Axiflavon are mainly due to its coupling to photons, and we summarize the status of  the relevant experiments in the usual  $(m_a, g_{a \gamma \gamma})$ plane in Fig.~\ref{plot}, where $g_{a \gamma \gamma} = |8/3 - 1.92| \, \alpha_{\rm em}/(2 \pi f_a)$. 
%
\section{Summary and Conclusions}

In summary, we have a proposed a $U(2)_F$ model of flavor with horizontal quantum numbers compatible with an $SU(5)$ GUT structure. The flavor symmetry $U(2)_F \overset{{\rm \tiny loc.}}{\simeq} SU(2)_F \times U(1)_F$ is spontaneously broken by two flavon fields $\phi$ and $\chi$, which transform as a doublet and singlet under $SU(2)_F$, respectively. Similarly, the three generations of SM fermions transform as ${\bf 2 +1 }$ of $SU(2)_F$, and there is a simple assignment of $U(1)_F$ quantum numbers
\begin{align}
X_{{10}_3} & = 0 \, , & X_{{10}_a} & = X_{{\overline{5}}_a} = X_{{\overline{5}}_3} = - X_\phi = - X_\chi =1  \, .
\label{conclus1}
\end{align}
The SM Yukawas arise from higher-dimensional operators made invariant under $U(2)_F$ by appropriate insertions of flavons, suppressed by the cut-off scale $\Lambda \gg v$. In this way the hierarchical structure of Yukawa matrices is explained by powers of two small parameters that control the breaking of $U(2)_F$, up to Wilson coefficients that are required to be ${\cal O}(1)$. The resulting Yukawa matrices in the quark and charged lepton sector have a simple structure with three texture zeros in the 1-1,1-3 and 3-1 entries, while the 1-2 entry is antisymmetric, see Eq.~\eqref{eq:yfapp}. The presence of these textures leads to accurate relations between CKM elements and masses, cf. Eq~\eqref{CKMpred}, which in contrast to the original $U(2)$ flavor models in Refs.~\cite{BarbieriU21, BarbieriU22} can be consistent with experimental data because of large rotations in the right-handed down quark sector. Indeed we have obtained a very good fit to fermion masses and mixings with coefficients that are  ${\cal O}(1)$ (all between 0.4 and 2), see Table~\ref{tab:bestFitsQuarkAndChargedLeptonsCP}.

\bigskip

We have then included the neutrino sector, which gives a consistent fit to experimental data only with Dirac neutrinos. To this extent, we have introduced three right-handed neutrinos (SM singlets), which also transform as  ${\bf 2 +1 }$ of $SU(2)_F$ and have equal charges under $U(1)_F$. The resulting structure of the Dirac mass matrix (cf. Eq.~\eqref{mnuD}) has again three texture zeros and only weak inter-generational hierarchies, thus predicting large mixing angles. The $U(1)_F$ charge of the singlets enters only in the overall suppression factor and can account for the smallness of neutrino Yukawas if taken to be $5 \div 6$.  The  combined fit to the complete fermion sector is viable only for neutrinos with normal mass hierarchy, and still shows a good performance with ${\cal O}(1)$ coefficients between roughly $1/3$ and $3$  (cf. Table~\ref{tab:bestFitsCombinedQuarkLeptonnu}). This fit determines all parameters in the neutrino sector, and thus gives predictions for the absolute neutrino mass scale and the related observables. Scanning over many good fits we have obtained a range for the sum of neutrino masses roughly given by $(58 \div 110)$ meV, while the prediction for the effective neutrino mass measured in $\beta$-decays is far below future experimental sensitivities. 

\bigskip

In order to have a consistent scenario with Majorana neutrinos, we have futhermore discussed an $D_6 \times U(1)$ variant of the $U(2)_F$ model, where the $SU(2)_F$ factor is replaced by a discrete $D_6$ subgroup. The charge assignment of fermions and spurions closely resembles the $U(2)_F$ structure, so that the effective Yukawa matrices in the quark and charged lepton sector are exactly the same as in the $U(2)_F$ case, up to a sign flip in the 1-2 entry that is largely irrelevant. This sign flip however allows for an unsuppressed 1-2 entry in the Weinberg operator, whose hierarchical structure follows directly from charges of the SM lepton doublets, and are to large extent independent of the charges of the heavy right-handed neutrinos (cf. Eq.~\eqref{mnuMesti}). Remarkably, the resulting structure automatically leads to an anarchic neutrino mass matrix, so that the $SU(5)$ structure connects large leptonic mixing angles to small mixing angles in the quark sector. Indeed, the parametric flavor suppression of up-, down-quark, charged lepton and neutrino masses follows the simple pattern 
\begin{align}
m_{\{ u, d, e, \nu \}} \sim \begin{pmatrix} 0 &   \eps^2 & 0 \\%
		\eps^2 & \eps^2 & \{ \eps, \eps^2, \eps, \eps^2 \}  \\%
		0  & \{ \eps, \eps, \eps^2, \eps^2 \} & \{ 1, \eps, \eps, \eps^2 \} \end{pmatrix} \, , %
\end{align}
where the mass scale is set by $v$ in the quark and charged lepton sector and $v^2/M$ in the neutrino sector. The difference between the fermion sectors just follows from the different $U(1)$ charge assignments for the third generation, see Eq.~\eqref{conclus1}. Although this model is more predictive than the Dirac case, since two $U(1)$ charges are replaced by a single mass scale $M$, we obtain an excellent fit all SM observables with ${\cal O}(1)$ coefficients between 0.4 and 2, see Table~\ref{tab:bestFitsCombinedQuarkLeptonnuM}. From this fit we can again predict the overall neutrino mass scales, and  as in the previous case only neutrinos with normal mass hierarchy are viable. Scanning over many good fits, we have obtained a slightly narrower range for the sum of neutrino masses roughly given by $(58 \div 78)$ meV, while again the predictions for the effective neutrino mass entering beta decay and neutrinoless double beta decay are far below future experimental sensitivities, see Table~\ref{rangeD6}. 

\bigskip

Finally we have discussed the various possibilities to test our models apart from the predictions in the neutrino sector. In general, sizable deviations in experimental observables from the SM require the existence of sufficiently light degrees of freedom. While there is no particular reason why the cutoff and its associated dynamics should be light, there is the natural possibility to solve the strong CP problem and account for DM through the Goldstone boson of the global $U(1)_F$ symmetry, which we refer to as the $U(2)$ Axiflavon. In contrast to the Axiflavon from a single Froggatt-Nielsen $U(1)_F$ symmetry~\cite{FN} as presented in Refs.~\cite{Axiflavon}, here the flavor-violating couplings of the axion are protected by the approximate $U(2)$ symmetry. Therefore, the $U(2)$ Axiflavon looks very much like a usual DFSZ/KSVZ axion, with the strongest constraint from WD cooling, which requires a sufficiently light axion $m_a < 14$ meV. Particularly interesting is the axion mass range where DM can be explained through the misalignment mechanism, implying axion masses around $(1 \div 40) \, \mu$eV, which corresponds to a cutoff scale of roughly  $(10^{13} \div 10^{15}) \GeV$. This range will be tested by future axion haloscope searches. 

\bigskip

The present model could be extended in several ways:  1) A more careful study of the neutrino sector might allow to pin down the predictions analytically, and it could be interesting to take a closer look to the type-I seesaw model, in particular its connection with Leptogenesis.  2) One could embed the model into a supersymmetric framework to address the hierarchy problem, possibly in connection with a full $SU(5)$ GUT, trying to relate GUT breaking scale, flavor breaking scale and the axion decay constant, similar to Ref.~\cite{Ringwald}. 3) Finally, it might be interesting to study possible UV completions and calculate the low-energy constraints from flavor-violating obervables on the new dynamics.

\section*{Acknowledgements}
We thank F.~Feruglio, J.~Lopez Pavon, A.~Ringwald, A.~Romanino and A.~Trautner for useful discussions and comments. ML acknowledges the support by the DFG-funded Doctoral School ''Karlsruhe School of Elementary and Astroparticle Physics: Science and Technology".

\newpage 

\appendix
\numberwithin{equation}{section}
\section{$D_3$ and $D_6$ Group Theory}   
\label{D3theory}
In this Appendix we provide some details about the structure of the dihedral groups $D_3$ and $D_6$ and fix the notation for constructing group invariants (see also Refs.~\cite{Dermisek:1999vy, Blum:2007jz,  Grimus:2011fk}).

The dihedral group $D_3$ is the symmetry group of an equilateral triangle and is isomorphic to $S_3$, the permutation group of three objects with order 6.  The group is generated by two elements $R$ and $S$, where $R$ is the rotation through $120^\circ$ and $S$ is  the reflection about one of the bisectors. Since $R^3 = S^2 = 1$ and $SR = R^2 S$, the six elements are $1, R, R^2, S, RS, SR$. \\%
$D_3$ has two one-dimensional representations $\mathbf{1}$, $\mathbf{1'}$ and one two-dimensional representation $\mathbf{2}$. The representation matrices for $R$ and $S$ can be chosen as in Table \ref{D3reps}.%
\begin{table}[h]%
	\centering%
	\label{tab:representationMatricesD3}%
	\begin{tabular}{ccc}%
		\toprule%
		Representation & $R$ & $S$\\%
		\midrule%
		$\mathbf{1}$ & \num{1} & \num{1}\\%
		$\mathbf{1'}$ & \num{1} & \num{-1}\\%
		$\mathbf{2}$ & $\begin{pmatrix}  e^{\frac{2 \pi i}{3}} & \\ & e^{\frac{- 2 \pi i}{3}}  \end{pmatrix}$ & $\begin{pmatrix} 0 & 1 \\ 1 & 0 \end{pmatrix} $\\%
		\bottomrule%
	\end{tabular}
		\caption{Representation matrices for $D_3$. \label{D3reps}}%
\end{table}
The tensor products of two one-dimensional representations decompose as follows:%
\begin{align}%
	\mathbf{1} \otimes \mathbf{1} &= \mathbf{1} \, , & 
	\mathbf{1} \otimes \mathbf{1'} &= \mathbf{1'}  \, , & 
	\mathbf{1'} \otimes \mathbf{1'} &= \mathbf{1} \, , \label{eq:D3Decomposition1PrimeTimes1Prime}%
\end{align}%
while for the product of two $\mathbf{2}$'s one gets%
\begin{equation}%
	\mathbf{2} \otimes \mathbf{2} = \mathbf{1} \oplus \mathbf{1'} \oplus \mathbf{2} \label{eq:D3Decomposition2Times2} \, .
\end{equation}%
For two doublets $\psi = \begin{pmatrix} \psi_1 \\ \psi_2 \end{pmatrix}$ and $\varphi = \begin{pmatrix} \varphi_1 \\ \varphi_2  \end{pmatrix}$ one finds %
\begin{align}%
	\left( \psi \otimes \varphi \right)_\mathbf{1} &= \psi_1 \varphi_2 + \psi_2 \varphi_1 \, , & %
	\left( \psi \otimes \varphi \right)_\mathbf{1'} &= \psi_1 \varphi_2 - \psi_2 \varphi_1 \, , & %
	\left( \psi \otimes \varphi \right)_\mathbf{2} &= \begin{pmatrix} \psi_2 \varphi_2 \\ \psi_1 \varphi_1 
	\end{pmatrix}  \, . \label{eq:D3Decomposition2Times22}%
\end{align}%
In the following we will use the simplified notation for singlet components (i.e. invariants)
\begin{align}
(\psi \cdot \varphi ) \equiv \left( \psi \otimes \varphi \right)_\mathbf{1} = \psi_1 \varphi_2 + \psi_2 \varphi_1 \, .
\end{align}
From a given doublet $\varphi$ one can construct another doublet  $\widetilde{\varphi} = \sigma^1 \varphi^\ast = \begin{pmatrix} \varphi_2^\ast \\ \varphi_1^\ast \end{pmatrix}$,  with  invariant%
\begin{equation}%
	(\widetilde{\varphi} \cdot \varphi) = \varphi_1^\ast \varphi_1 + \varphi_2^\ast \varphi_2  \, .\label{eq:D3Decomposition2TildeTimes2}%
\end{equation}%
Note that because of Eq.~\eqref{eq:D3Decomposition2Times2} any product of doublets contain at least one singlet. For three doublets it is given by
\begin{equation}%
	\left( \psi \cdot \varphi \cdot \chi \right) = \psi_1 \varphi_1 \chi_1 + \psi_2 \varphi_2 \chi_2 \label{eq:D3SingletFrom222} \, ,
\end{equation}%
while there are three different singlets in the product of four doublets, which we define as 
\begin{align}
\left( \psi \otimes \varphi \otimes \chi \otimes \eta \right)_\mathbf{1} =
\begin{cases}
	\psi_1 \varphi_2 \chi_1 \eta_2 + \psi_2 \varphi_1 \chi_2 \eta_1  \\%
	\psi_1 \varphi_2 \chi_2 \eta_1 + \psi_2 \varphi_1 \chi_1 \eta_2  \\%
	\psi_1 \varphi_1 \chi_2 \eta_2 + \psi_2 \varphi_2 \chi_1 \eta_1 %
\end{cases} \, .
\end{align}
For the case of $\psi = \varphi$ and $\chi = \eta$  there are just two invariants for which we use the notation:
\begin{align}
\left( \psi \otimes \psi \otimes \chi \otimes \chi \right)_\mathbf{1} =
\begin{cases}
(\psi \cdot \psi) (\chi  \cdot \chi) & \equiv 4 \,  \psi_1 \psi_2 \chi_1 \chi_2  \\ (\psi \cdot \psi \cdot \chi \cdot \chi ) & \equiv \psi_1^2 \chi_2^2 + \psi_2^2 \chi_1^2 
\end{cases} \, .
\end{align}
Finally we turn to the dihedral group $D_6$ which is the symmetry group of  regular hexagon. It is isomorphic to $D_3 \times Z_2$, and therefore inherits the group theoretical structure discussed above, except that each representation carries an additional $Z_2$ charge, which is conserved in tensor decompositions. Thus, we have four one-dimensional representations ${\bf 1_+, 1_-, 1^\prime_+, 1^\prime_-}$ (where ${\bf 1_+}$ denotes the total singlet) and two two-dimensional representations ${\bf 2_+, 2_-}$. The decompositions of these representations follow from the $D_3$ ones, for example we have
\begin{align}
	\mathbf{2}_- \otimes \mathbf{2}_- & = \mathbf{1}_+ \oplus \mathbf{1'}_+ \oplus \mathbf{2}_+  \, , & 
	\mathbf{2}_+ \otimes \mathbf{2}_- & = \mathbf{1}_- \oplus \mathbf{1'}_- \oplus \mathbf{2}_-  \, .
\end{align}
Therefore in $D_6$ the tensor product $({
\bf 2}_- \otimes {\bf 2}_- \otimes {\bf 2}_-)$ does not contain a singlet.  
\section{Fit Results}

\begin{table}[H]
    \centering
	    \begin{tabular}{|c||cccc|cc|}
        \hline
        Parameter & {QL$\nu_D$-1} & {QL$\nu_D$-2} & {QL$\nu_D$-3} & {QL$\nu_D$-4} & {QL$\nu_M$-1}  & {QL$\nu_M$-2} \\
        \hline
		$\lambda^u_{12}$ & 0.902 & 0.843 & 3.831 & 1.162 & -1.633 & -1.176\\
		$\lambda^u_{22}$ & 1.187 & -1.047 & 1.859 & 1.148 & 1.339 & 1.112\\
		$\lambda^u_{23}$ & 2.222 & -2.175 & -2.138 & -1.799 & 2.127 & 1.925\\
		$\lambda^u_{32}$ & -1.103 & -1.419 & 1.511 & 2.422 & 1.196 & 1.615 \\
		$\lambda^u_{33}$ & 0.787 & 0.779 & -0.787 & 0.786 & 0.787 & 0.785 \\
		$\delta_{33}$ & -0.640 & -0.720 & -3.948 & -1.097 & -3.837 & -3.988 \\
		\hline
		$\lambda^d_{12}$ & 0.479 & -0.479 & 2.165 & 2.173 & -0.888 & 0.976 \\
		$\lambda^d_{22}$ & -1.000 & -1.156 & -1.075 & -0.972 & -0.973 & 0.976 \\
		$\lambda^d_{23}$ & 0.913 & -0.786 & -1.304 & -1.155 & 1.073 & 0.985\\
		$\lambda^d_{32}$ & -0.355 & 0.401 & 0.414 & 0.423 & 0.365 & -0.394 \\
		$\lambda^d_{33}$ & 0.665 & 0.651 & 1.394 & 1.497  & -0.902 & -0.948 \\
		\hline
		$\lambda^\ell_{12}$ & 0.402 & -0.376 & -1.752 & -1.758 & -0.801 & 0.856 \\
		$\lambda^\ell_{22}$ & 0.987 & -1.134 & 1.821 & 2.052 & 1.306 & 1.497 \\
		$\lambda^\ell_{23}$ & 0.343 & 0.381 & 0.393 & -0.414 & -0.368 & 0.391 \\
		$\lambda^\ell_{32}$ & -0.992 & -1.132 & 1.175 & 1.193 & -1.198 & 1.294 \\
		$\lambda^\ell_{33}$ & 0.432 & -0.399 & -0.945 & 0.992 & -0.503 & -0.536 \\
		\hline
		$\lambda^\nu_{12}$ & 0.882 & -1.416 & 0.938 & 1.006 & 2.130 & -1.873 \\
		$\lambda^\nu_{22}$ & -0.994 & -1.303 & 0.325 & 0.398 & -0.844 & -0.760 \\
		$\lambda^\nu_{23}$ & -2.588 & -1.074 & -1.505 & 1.681 & 1.137 & -1.078 \\
		$\lambda^\nu_{32}$ & 1.065 & -0.704 & 0.601 & 0.680 & $\shortparallel$ & $\shortparallel$ \\
		$\lambda^\nu_{33}$ & 0.952 & -1.572 & -0.890 & 0.891 & -0.489 & -0.655 \\
		\hline
		$X^\mathrm{N}_a$ & {6} & {6} & {5} & {5}  &   &   \\
		$X^\mathrm{N}_3$ & {6} & {6} & {5} & {5} &   &  \\
		$v/M \times 10^9$ &   &  &   &   & -0.421 & -0.520 \\
		\hline
		$\varepsilon_\phi$ & 0.026 & 0.024 & 0.022 & 0.021  & 0.025 & 0.024 \\
		$\varepsilon_\chi$ & 0.012 & 0.013 & 0.006 & 0.006  & 0.009 & 0.009 \\
        \hline
    \end{tabular}
    \caption{Fit parameters for Dirac ($SU(2) \times U(1)$ Model) and Majorana neutrinos ($D_6 \times U(1)$ Model). The parameters are defined in Eqs.~\eqref{eq:yfapp} and \eqref{mnuD}, and Eq.~\eqref{eq:yallD6}, respectively. 
	\label{tab:fitResults}}
\end{table}
Below we also provide the finetuning and pulls of the fit. For each observable $O_i = \{ y_u, y_d, \ldots\}$ we define the tuning $\Delta_i$ and the pull $P_i$ as
\begin{align}
\Delta_i & = {\rm max}_j \left| \frac{\partial \log O_i}{\partial \log p_j} \right| \, , & 
P_i & = \frac{O_i^{\rm fit} - O_i^{\rm exp}}{\sigma_i^{\rm exp}} \, , 
\end{align}
where $p_j = \{ \lambda^{u,d,\ell,\nu}_{ij} , \eps_\phi , \eps_\chi, M\}$ are the fit parameters. For the sake of brevity, we restrict to Fit 3 and 4 in the Dirac case, the other two fits give similar results. As can be seen from Tables~\ref{FTD} and \ref{FTM}, the tuning of the observables is quite low, at most 10\% for the Dirac case and about 20\% in the Majorana case. As expected from the $\chi^2$ value, the pulls are small and are dominated by the quark Yukawas (and in the Majorana case also by the PMNS mixing angles). 
\begin{table}[H]
	\centering
		\small
	\begin{tabular}{|c||cc|cc|}
		\hline
		Observable &  $\Delta$(QL$\nu_D$-3) & Pull(QL$\nu_D$-3)  &  $\Delta$(QL$\nu_D$-4)  & Pull(QL$\nu_D$-4)  \\
		\hline
 		$y_{u}$              &  4   & 0.0  &  4   & -2.3 \\
 		$y_{c}$             &  2   & -0.1  &  2   & -1.2 \\
 		$y_{t}$            &  1   & -0.0 &  1   & -0.1  \\
 		$y_{d}$                  &  3.8 & 0.9 &  3.8 & 2.1  \\
 		$y_{s}$             &  1.4 & -0.2  &  1.3 & -2.3 \\
 		$y_{b}$             &  0.6 & -0.0 &  0.5 & 0.1  \\
 		$y_{e}$               &  4   & -0.0 &  4   & -0.0 \\
 		$y_{\mu}$            &  1.3 & 0.0  &  1.3 & -0.0 \\
 		$y_{\tau}$             &  0.7 & 0.0  &  0.7 & 0.0 \\
 		$\theta^\mathrm{CKM}_{13}$  &  1.2 & -0.1  &  1.1 & -0.1 \\
 		$\theta^\mathrm{CKM}_{12}$  &  1.8 & -0.1 &  1.7 & -0.2 \\
 		$\theta^\mathrm{CKM}_{23}$ &  3.8 & -0.0  &  1.1 & 0.0  \\
 		$\delta_\mathrm{CP}$        &  4.8 & -0.0 &  0.8 & 0.6 \\
 		$\Delta m^2_{21}$            &  11.3 & -0.0 & 11.4 & -0.3 \\
 		$\Delta m^2_{31}$            &  9.3 & 0.0  &   9.3 & 0.3 \\
 		$\theta^\mathrm{PMNS}_{13}$ &  1.3 & 0.1 &  1.3 & 0.6 \\
 		$\theta^\mathrm{PMNS}_{12}$ &   1.1 & 0.0  &  1.1 & -0.5 \\
 		$\theta^\mathrm{PMNS}_{23}$ & 0.7 & 0.0   &  0.7 & 0.7  \\
		\hline
	\end{tabular}
	\caption{Fine-tuning and pulls for the observables of fit QL$\nu_D$-3 and fit QL$\nu_D$-4. \label{FTD}}
\end{table}

\begin{table}[H]
	\centering
	\small
	\begin{tabular}{|c||cc|cc|}
		\hline
		Observable &  $\Delta$(QL$\nu_M$-1) & Pull(QL$\nu_M$-1) &  $\Delta$(QL$\nu_M$-2) & Pull(QL$\nu_M$-2) \\
		\hline
 		$y_{u}$            & 4           & 0.1   & 4           & -1.3 \\
 		$y_{c}$            & 2           & -0.1  & 2           & -1.4 \\
 		$y_{t}$                  & 1           & -0.0  & 1           & -0.3 \\
 		$y_{d}$             & 3.8         & 0.7 & 3.8         & 1.6  \\
 		$y_{s}$            & 1.6         & -0.1 & 1.5         & -1.4  \\
 		$y_{b}$            & 0.6         & -0.0 & 0.6         & 0.8   \\
 		$y_{e}$            & 4           & -0.0  & 4           & -0.0  \\
 		$y_{\mu}$           & 1.2         & -0.0  & 1.2         & -0.0  \\
 		$y_{\tau}$             & 0.8         & 0.0  & 0.8         & 0.1  \\
 		$\theta^\mathrm{CKM}_{13}$    & 1.3         & -0.1  & 1.2         & 0.4  \\
 		$\theta^\mathrm{CKM}_{12}$    & 1.8         & -0.0& 1.7         & -0.1  \\
 		$\theta^\mathrm{CKM}_{23}$& 4.5         & -0.0 & 4           & -0.1 \\ 
 		$\delta_\mathrm{CP}$        &   5.1         & -0.1& 3.9         & 0.0 \\
  		$\Delta m^2_{21}$          & 4.7         & 0.1  & 4.7         & 0.5 \\
 		$\Delta m^2_{31}$              & 3.1         & -0.0  & 3           & -0.2 \\
 		$\theta^\mathrm{PMNS}_{13}$  &  1.4         & -0.1 & 1.4         & -1.4 \\
 		$\theta^\mathrm{PMNS}_{12}$  &  1.1         & 0.2 & 1.1         & 2.3  \\
 		$\theta^\mathrm{PMNS}_{23}$  &  0.8         & 0.4 & 0.8         & 1.6   \\
 		\hline
	\end{tabular}
	\caption{Fine-tuning and pulls for the observables of fit QL$\nu_M$-1 and fit QL$\nu_M$-2.  \label{FTM}}
	\end{table}

\section{Scalar Potential }
\label{Vscal}
In this section we consider an explicit scalar potential that generates the VEVs we have assumed in Section~\ref{D6}, serving merely as a proof of existence. In particular, this potential should be reassessed in a UV complete setup, possibly in connection with a supersymmetric $SU(5)$ GUT. 

In addition to the scalars $\phi$ and $\chi$ we need to introduce a new (SM singlet) scalar $\psi$ in order to break the $U(1)$ symmetries in the scalar potential to a single continuous global symmetry that can be identified with $U(1)_F$. The transformation properties under $D_6 [U(1)_F]$ are
\begin{align}
\phi & = 2_- [-1] \, , & \chi & = 1_+ [-1] \, , & \psi & = 1_- [+1] \, , 
\end{align}
and the most general, renormalizable scalar potential for these fields is given by\footnote{We do not include the SM Higgs, because its backreaction on the flavon potential is negligible as the flavon VEVs are much larger than the electroweak scale. In turn, the flavons will generate a large mass term for the Higgs, which is just the usual hierarchy problem that we do not address here.}
\begin{align}
V_{\rm scal} & = m_\chi^2 |\chi|^2 +   \left(  m_\phi^2 + \kappa_\chi |\chi|^2 + \kappa_\psi |\psi|^2 \right)  (  \tilde{\phi} \cdot \phi )+ m_\psi^2 |\psi|^2  \nonumber \\
& + \frac{\lambda_1}{4}  (\tilde{\phi} \cdot \tilde{\phi}) (\phi \cdot  \phi) + \frac{\lambda_2}{2}  (\tilde{\phi} \cdot \tilde{\phi} \cdot \phi \cdot \phi   ) + \lambda_3 |\chi^2| |\psi|^2 + \frac{\lambda_\chi}{2} |\chi|^4 + \lambda_\psi |\psi|^4  \nonumber \\
& + \left[ \frac{\kappa_1}{2} \psi \psi \left( \phi \cdot \phi \right) + \frac{\kappa_2}{2} \chi^* \chi^* \left( \phi \cdot \phi \right) + \frac{1}{2} \lambda_{\chi \psi}  \psi \psi \chi \chi + \rho \psi ( \tilde{\phi} \cdot \phi \cdot \phi  ) + {\rm h.c.} \right] \, ,
\end{align}
where the $D_6$ singlet contractions are explained in Appendix~\ref{D3theory} and we take $\kappa_1, \kappa_2, \lambda_{\chi \psi}$ and $\rho$ to be real. The ground state of this potential is most easily studied in the limit when 
\begin{align} 
\rho & \ll 1, & \kappa_2 & \ll1 \, , & \lambda_{\chi \psi} & \ll1 \, .
\end{align}  
For a suitable range of parameters (see below), one can easily show that the ground state at leading order in $\rho$ and $\kappa_2$ is given by 
\begin{align}
v_1^2 & = \frac{\lambda_\chi m_\phi^2 - \kappa_\chi m_\chi^2 }{\kappa_\chi^2 - \lambda_2 \lambda_\chi} \, , & 
v_\chi^2 & = \frac{\lambda_2 m_\chi^2 - \kappa_\chi m_\phi^2 }{\kappa_\chi^2 - \lambda_2 \lambda_\chi} \, .
\label{v1vchi}
\end{align}
There is a symmetry exchanging $\phi_1 \leftrightarrow \phi_2$ in the potential, which are connected by a $D_6$ transformation that we can use to assume the large VEV in the $\phi_1$ direction without loss of generality. The VEVs of $v_2$ and $v_\psi$ only arise at ${\cal O}(\kappa)$ and  ${\cal O}(\kappa \rho)$, respectively:
\begin{align}
v_2^2 & =  \frac{\kappa_2^2 v_\chi^2}{(\lambda_2 - \lambda_1)^2} \frac{v_\chi^2}{v_1^2} \, , & v_\psi^2 & = \frac{\kappa_2^2 \rho^2 v_1^2}{(\lambda_2 - \lambda_1)^2} \left( \frac{v_\chi ^2  (\kappa_\chi^2 - \lambda_2 \lambda_\chi)}{\tilde{m}^2} \right)^2 \, , 
\label{v2vpsi}
\end{align}
with the shorthand notation
\begin{align}
\tilde{m}^2 & = \kappa_\chi (\kappa_\chi m_\psi^2 - \kappa_\psi m_\chi^2 - \lambda_3 m_\phi^2  ) + \kappa_\psi \lambda_\chi m_\phi^2 + \lambda_2 ( \lambda_3 m_\chi^2 -  \lambda_\chi m_\psi^2 ) \, .
\end{align}
In order to suppress the VEVs of $\phi_2$ and $\psi$ sufficiently, i.e. to ensure the validity of e.g. Eq.~\eqref{eq:yfapp}, we need roughly $v_2/v_1 \sim v_2/v_\chi \sim \kappa_2 \lesssim \eps_\chi^2 \sim 10^{-4}$. Such a small coupling is technically natural, since in the limit of $\kappa_2 \to 0, \lambda_{\chi \psi} \to 0$ (or $\rho \to 0$) the Lagrangian acquires a larger symmetry. This can be seen from spelling out the third line of the potential explicitly:
\begin{align}
V_{\rm scal} & \supset \kappa_1 \psi^2 \phi_1 \phi_2 + \kappa_2 \chi^* \chi^* \phi_1 \phi_2 + \lambda_{\chi \psi} \chi^2 \psi^2 + \rho \psi \left( \phi_1^2 \phi_2^*+ \phi_2^2 \phi_1^* \right) + {\rm h.c.}
\end{align}
Indeed, this part only breaks the additional $U(1)^3$ symmetry of the scalar kinetic terms  (besides the remaining $U(1)_F$) if $\rho \ne 0$ and  $\kappa \ne 0$ or $\lambda_{\chi \psi} \ne0$. 

This observation is also crucial to understand why the additional field $\psi$ is needed: its coupling $\rho$ is the only parameter that breaks the $U(1)$ symmetry under which $\chi$ is neutral and $\phi_1$ and $\phi_2$ carry opposite charges. Moreover, it makes clear that we expect (in addition to the massless $U(1)_F$ Goldstone) a very light pseudoscalar in the spectrum whose mass is suppressed by the small couplings $ \kappa_2, \lambda_{\chi \psi}$ and $\rho$. 

After these analytical considerations we finally provide a numerical example, taking the following set of parameters:
\begin{gather}
m_\phi^2 = - 2 m^2 \, , m_\chi^2  = - 3/10 m^2\, ,  m_\psi^2 = 2 m^2 \, , \lambda_1  = 1 \, , \lambda_2 = 1/9 \, , \lambda_\chi = 1 \, , \kappa_\chi = -1/8 \nonumber \\
\lambda_3  = 2/3 \, , \lambda_{\chi \psi}  = - 1/20 \, , \kappa_1  = -1/3 \, , \kappa_\psi = 7/10 \, , \rho = - 1/20 \, , \lambda_\psi = 9/10 \, , \kappa_2 = 1/2000 
\end{gather}
The absolute minimum in the potential can be calculated numerically, and agrees very well with the above approximate results in Eq.~\eqref{v1vchi} and Eq.~\eqref{v2vpsi}. The VEVs are given by
\begin{align}
v_1 & = 4.6  m \, , & v_2 & = - 3.6 \cdot 10^{-4} \, m \, , & v_\chi & = 1.7 m \, , & v_\psi & = - 2.1 \cdot 10^{-5} \, m \, ,
\end{align}
and therefore
\begin{align}
\eps_\phi & = 0.024 \left( \frac{190 m}{ \Lambda} \right) \, , &
\eps_\chi & = 0.009 \left( \frac{190 m}{ \Lambda} \right) \, .
\end{align}
Finally, the scalar mass spectrum is given by one massless Goldstone, 6 massive scalars with masses $\{6.3, 6.3, 6.0, 6.0, 3.9, 2.6\} m$ and a light scalar with mass $2.9 \cdot 10^{-5} m$. Using the lower bound on $\Lambda$ from Section~\ref{axionpheno} (corresponding to an axion mass $m_a \lesssim 14$ meV), we get a lower bound on $m$ roughly given by $m \sim  \Lambda/190 > 5 \cdot 10^8 \GeV $, so the light scalar has a mass $ \gtrsim 15 \TeV$.

We finally comment on the small value of $\kappa_2 = 1/2000$ used in the benchmark point. As it is clear from Eq.~\eqref{v2vpsi}, small $\kappa_2$ ensures the approximate alignment of the doublet VEV along $v_1$.  This small value is technically natural within the benchmark point, since the renormalization group equation for $\kappa_2$ is of the form $d \kappa_2/dt \sim \lambda_{\chi \psi}^* \kappa_1/16 \pi^2$, so radiative corrections to $\kappa_2$ are under control. Within the context of a supersymmetric UV completion there might be a more natural possibility to ensure the VEV alignment of $\phi$. 

\bibliographystyle{JHEP} 
 \bibliography{Mybibliography}

\end{document}